\documentclass[fleqn,usenatbib]{mnras}
\usepackage{newtxtext,newtxmath}
\usepackage{comment}
\usepackage{multirow}
\usepackage{subcaption}
\usepackage{longtable}
\usepackage{supertabular}
\usepackage{anyfontsize}
\usepackage{placeins}
\usepackage{float}
\usepackage{hyperref}
\usepackage{amsmath} 

\usepackage[T1]{fontenc}
\usepackage{booktabs}
\usepackage{siunitx}
\usepackage{xspace} 

\DeclareRobustCommand{\VAN}[3]{#2}
\let\VANthebibliography\thebibliography
\def\thebibliography{\DeclareRobustCommand{\VAN}[3]{##3}\VANthebibliography}

\usepackage{graphicx}	
\usepackage{amsmath}
\usepackage{amsbsy}
\usepackage{adjustbox} 
\usepackage{subcaption}
\usepackage{balance}
\usepackage{caption}

\newcommand{\kms}{\ensuremath{\mathrm{km\,s^{-1}}}\xspace}

\def\equationautorefname~#1\null{Eq.~#1\null}

\newcommand{\orcidsymb}[2]{\href{http://orcid.org/#2}{#1\adjustbox{trim={-.15\width} {0\height} {-.15\width} {0\height},clip}{\includegraphics[height=10pt]{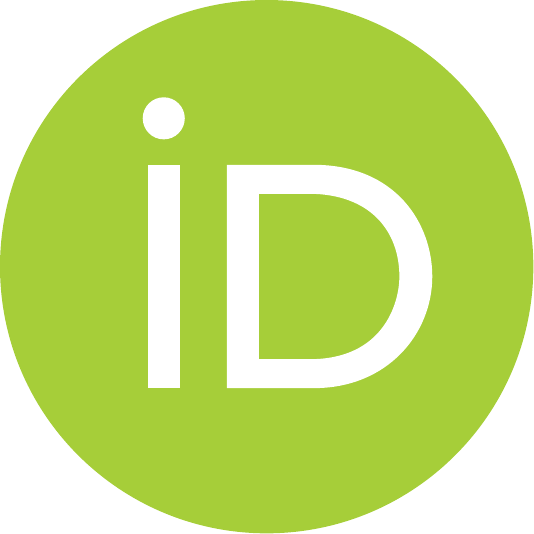}}}}

\title[GS-9209: NIRSpec/IFU resolved stellar kinematics]{When relics were made: vigorous stellar rotation and low dark matter content in the massive ultra-compact galaxy GS-9209 at z=4.66}

\author[Robert Pascalau et al.]{
\parbox{\textwidth}{
\orcidsymb{Robert G. Pascalau,}{0000-0001-9820-5773}$^{1,2}$\thanks{E-mail: rgp34@cam.ac.uk} 
\orcidsymb{Francesco D'Eugenio,}{0000-0003-2388-8172}$^{1,2}$\thanks{E-mail: fdeugenio@gmail.com}
\orcidsymb{Sandro Tacchella,}{0000-0002-8224-4505}$^{1,2}$
\orcidsymb{Roberto Maiolino,}{0000-0002-4985-3819}$^{1,2,3}$
\orcidsymb{Michele Cappellari,}{0000-0002-1283-8420}$^{4}$
\orcidsymb{Qiao Duan,}{0009-0009-8105-4564}$^{1,2}$
\orcidsymb{Claudia del P. Lagos,}{0000-0003-3021-8564}$^{5,6}$
\orcidsymb{Andrew J. Bunker,}{0000-0002-8651-9879}$^{4}$
\orcidsymb{Gareth C. Jones,}{0000-0002-0267-9024}$^{1,2,4}$
\orcidsymb{Jan Scholtz,}{0000-0001-6010-6809}$^{1,2}$
\orcidsymb{Hannah {\"U}bler,}{0000-0003-4891-0794}$^{7}$
\orcidsymb{Giovanni Cresci,}{0000-0002-5281-1417}$^{8}$
\orcidsymb{Santiago Arribas,}{0000-0001-7997-1640}$^{9}$
\orcidsymb{Michele Perna,}{0000-0002-0362-5941}$^{9}$
\orcidsymb{Arjen van der Wel,}{0000-0002-5027-0135}$^{10}$
\orcidsymb{A. Lola Danhaive,}{0000-0002-9708-9958}$^{1,2}$
\orcidsymb{William McClymont,}{0009-0009-5565-3790}$^{1,2}$
\orcidsymb{Christina C. Williams,}{0000-0002-2380-9801}$^{11,12}$
\orcidsymb{Anna de Graaff,}{0000-0002-2380-9801}$^{13}$
\orcidsymb{Akash Vani,}{0000-0002-3370-8086}$^{14,15}$
\orcidsymb{Michael V. Maseda,}{0000-0003-0695-4414}$^{16}$
\orcidsymb{Adam C. Carnall,}{0000-0002-1482-5818}$^{17}$
\orcidsymb{St\'ephane Charlot,}{0000-0003-3458-2275}$^{18}$
\orcidsymb{Stefano Carniani,}{0000-0002-6719-380X}$^{19}$
\orcidsymb{Tze P. Goh,}{0000-0001-8383-2299}$^{1,2}$
\orcidsymb{Zhiyuan Ji,}{0000-0001-7673-2257}$^{12}$ and
\orcidsymb{Pablo P\'erez Gonz\'alez}{0009-0009-7829-046X}$^{9}$
}
\vspace{0.25cm}
\\
Affiliations are listed at the end of the paper.
}
\date{\vspace{-11ex}}

\pubyear{2025}

\begin{document}
\label{firstpage}
\pagerange{\pageref{firstpage}--\pageref{lastpage}}
\maketitle


\begin{abstract}




JWST uncovered a large number of massive quiescent galaxies (MQGs) at $z>3$, which theoretical models struggle to reproduce. Explaining the number density of such objects requires extremely high conversion efficiency of baryons into stars in early dark matter halos. Using stellar kinematics, we can investigate the processes shaping the mass assembly histories of MQGs. We present high-resolution JWST/NIRSpec integral field spectroscopy of GS-9209, a massive, compact quiescent galaxy at $z=4.66$ ($\log \left (M_{\ast}/M_{\odot} \right) = 10.52 \pm 0.06 $, $R_{\rm eff} = 220 \pm 20$ pc). Full spectral fitting of the spatially resolved stellar continuum reveals a clear rotational pattern, yielding a spin parameter of $\lambda_{2R_{\rm eff}} = 0.85 \pm 0.10$. 
This study suggests that at least a fraction of the earliest quiescent galaxies were fast rotators and that quenching was a dynamically gentle process, preserving the stellar disc even in highly compact objects. Using Jeans anisotropic modelling and
assuming a NFW profile, we measure a dark matter fraction of $f_{\rm DM} \left (<2 R_{\rm eff} \right ) = 14.5^{+6.0}_{-4.2} \%$. Our findings use stellar kinematics to confirm the massive nature of early quiescent galaxies, previously inferred from stellar population modelling. We suggest that GS-9209 has a similar structure to low-redshift `relic' galaxies. However, unlike relic galaxies which have bottom-heavy initial mass functions (IMF), the dynamically inferred stellar mass-to-light ratio of GS-9209 is consistent with a Milky-Way like IMF. The kinematical properties of GS-9209 are different from those of $z<1$ early-type galaxies and more similar to those of recently quenched post-starburst galaxies at $z>2$.

\end{abstract}

\section{Introduction}
\label{sec:introduction}
Galaxies' star-formation histories (SFH) and observed stellar populations are fundamental to our understanding of galaxy evolution. The star formation activity of a galaxy is closely related to its morphology \citep[as it was revealed by studies on $0<z<2.5$ galaxies e.g.,][]{brennan2017,dimauro+2022}. A further link exists between galactic morphologies and kinematics with the classical paradigm splitting the galaxies population into rotationally supported flattened discs and dispersion-supported spheroids \citep{emsellem2007_lambda_def,atlas3_emsellem2011,vandesande2018,graham_sdss4,thob2019morpho_kin}. Such classifications seem to be valid up to a redshift of $z \sim 3$ \citep[e.g.,][]{brennan2017} but it is unclear whether they hold true for higher-$z$ galaxies \citep{abraham1999}. 

The question of how different types of galaxies are born, evolve, and end up being ``red and dead'' has been central in our quest to understand galactic evolution. Since the star-formation rate (SFR) is a key property of any galaxy, it goes without saying that quenching also has a dramatic impact on the evolution of both individual systems and the galaxy population as a whole. There is a wide variety of quenching mechanisms at play, depending on galaxy's stellar mass, kinematics and on its environment \citep{Man_2018,bluck2019quenching,bluck2020quenching,Cortese2021}. A few examples include: removal of cold molecular gas via outflows driven by active galactic nuclei (AGN, e.g. \citealt{fluetsch2019,rev2020outflows}), ram pressure stripping (the main mechanism in the case of low-mass galaxies, \citealt{gunn_gott_RPS}) or strangulation in the case of more massive galaxies (a complete halt of the galaxy's pristine gas inflows, \citealt{strangulation}). The latter is often the result of the halo heating due to supermassive black hole (SMBH) feedback \citep{agnfeedback98,bower2006,croton2006,lagos+2008,haloheating} but the gas might also get heated when falling into the dark matter halos' gravitational potential (virial shock heating, \citealt{white_rees1978,keres2005,dekel2006}). Star formation quenching is an important process in the evolution of galaxies, having a crucial effect on their subsequent morphologies. Furthermore, there is a strong correlation between a galaxy's average stellar population age and its morphology \citep{blanton2009,somme2015} or kinematics \citep{croom+24}.

Since the launch of James Webb Space Telescope \citep[\it JWST;][]{JWST_new,rigby+23} the sample of high-z massive quiescent galaxies (MQGs) has seen a dramatic increase in the number of studies on confirmed spectroscopic candidates in the redshift range $3<z<5$  \citep{carnall_CEERS_abundance2023,CarnallQuenching,carnall2023excels,Valentino_2023_nquench,rubiesEGS49,deugenio+2024,nanayakkara_quenched_pop,perez2024jekyll,setton2024_quench,baker2025} even going as early as $z \sim 7.3$ \citep{rubies73quenched2025}. These observations imply much higher number densities of massive quiescent galaxies compared with the predictions from numerical simulations or semi-analytic models (e.g. \textit{TNG300} - \citealt{hartley2023_sims}, \textit{Magneticum Pathfinder} - \citealt{blowing_out}, \textit{FLARES} - \citealt{lovell2023_sims}, \textit{SHARK v2.0} - \citealt{Lagos2024_sims}, \textit{FLAMINGO} - \citealt{baker2025},  \textit{L-Galaxies} - \citealt{Vani+25}). \citet{lagos+25} analysed six large cosmological simulations (3 semi-analytic models $-$ \textsc{GAEA}, \textsc{GALFORM} and \textsc{SHARK} $-$ and 3 full hydro-dynamical simulations: \textsc{SIMBA}, \textsc{EAGLE} and \textsc{IllustrisTNG}) and showed that these simulations predict that the number densities of $z \sim 4.5$ MQGs ($M_{\ast} > 10^{10.5} \ \rm M_{\odot}$, sSFR < 0.1 $\rm Gyr^{-1}$) are smaller than measured from {\it JWST} observations, the difference ranging from 1.5 to 3 dex depending on the simulation. Furthermore, it was demonstrated that the simulations are unable to reproduce the observed stellar mass functions of quiescent galaxies at $2<z<5$, as demonstrated by \citet{baker2025simulations}.

Even more puzzlingly, high redshift MQGs appear to have assembled large stellar masses ($M_{\ast} > 10^{10} \ \rm M_{\odot}$) during intense star formation episodes (up to $2000 \ M_{\odot} \rm \ yr^{-1}$ or more, \citealt{forrest2020quench2}) and then quench extremely rapidly in comparison to their local counterparts \citep{2_from_PSB,3_from_PSB}. Many observational studies have focused on the SFHs of these galaxies, but the processes governing their mass assembly history and quenching remain covered in mystery. MQGs pose at least three major challenges for current models of galaxy formation and evolution, resulting from: 1) their unexpected high number densities at high redshift (e.g. \citealt{CarnallQuenching,Valentino_2023_nquench,alberts+24,long2024_quench,zhang2025degraaff}), 2) their high stellar mass (possibly related to the high efficiency of baryon conversion into stars in high redshift massive halos, e.g. \citealt{dekel+23}) and 3) the rapid quenching mechanism. Within the standard paradigm of galaxy formation, the star formation and quenching timescales of massive quiescent galaxies imply the existence of impossibly massive halos at early times (e.g. \citealt{ZFUDS7329,turner2024oct,rubies73quenched2025}).

Continued star formation from the available cold gas would be more problematic, giving rise to impossibly massive galaxies, which would be inconsistent with both observations \citep{bower2012,harrison2018} and with the low redshift stellar mass function extrapolation \citep{mcleod2021}. Therefore, another issue to be discussed is the fast quenching timescale of high-$z$ MQGs given the early stage of the Universe when dark matter halos are accreting gas at extremely high rates \citep{sandro_tacchella2018}. These conditions point towards the possible responsible quenching mechanism being AGN ejective feedback \citep{agnfeedback98,dimatteo2005,maiolino2012agn_feedback,belli2024,xie2024,lim2025} as significantly more violent and more rapid than other quenching mechanisms \citep{sandro_tacchella2022}. In fact, about 50\% of MQGs at $1.5<z<4.5$ are AGN hosts \citep{bugiani2025agn,baker2025}. Furthermore, these claims of AGN feedback being responsible for quenching high-z MQGs are backed-up by theoretical studies such as \citet{piotrowska2022} or \citet{lagos+25}. Other theoretical works discuss possible alternative processes leading to a rapid gas consumption in MQGs, such as violent disc instabilities or feedback free star formation \citep{dekel+23,Vani+25}.

Indeed, many previous studies on MQGs show them to have low amounts of cold gas, the fuel star formation \citep[$f_{\rm gas} = M_{\rm gas}/\left(M_{\rm gas}+M_{\ast}\right)<20\%$; e.g.,][]{belli_coldgas,whittaker2021_coldgas,williams2021_coldgas,woodrum_coldgas,jan_net0}. However, other studies confirm the presence of non-negligible cold molecular gas reservoir that MQGs are unable to use for star formation. This is seen at various redshifts: $z \sim 0$ \citep{rowlands2015_coldgas} and $z \sim 1$ \citep{bezanson2022_coldgas}. The latter study reveals the extremely short cold gas depletion timescales ($\sim 100-200$ Myr) of recently quenched galaxies, interpreting this result as a possible evidence of AGN feedback. These studies highlight the contrast between the population of old quiescent galaxies and the ones who stopped forming stars more recently (post-starburst galaxies, PSBs). A similar classification of high redshift MQGs has emerged from studies of their star formation histories \citep{minjungpark2024_outflows}. 
In contrast to PSBs, the cold gas fractions of galaxies quenched long before the observation time ($>1$ Gyr) is more difficult to interpret. Using a sample of $z\sim0.7$ MQGs, \citet{woodrum_coldgas} find evidence for post-quenching rejuvenation effects driven by gas-rich minor mergers. However, \citet{suess+2023} found that the less massive, but more highly star-forming satellites of MQGs, although they do not necessarily bring substantial amounts of cold gas, they can contribute (via minor mergers) to up to $30\%$ of the central galaxies stellar mass.

In this paper, we present high-resolution $R \sim 2700$ NIRSpec integral-field-unit (IFU) observations of the massive quenched galaxy GS-9209 at redshift $z=4.66$, whose SFH and various structural and morphological properties (size, central BH, AGN properties, stellar and dust SED) have been studied in depth in \citet{CarnallQuenching} and \citet{paper_with_dust}. With the unique ability of NIRSpec/IFU to provide both complete spatial and spectral information, we focus here on the kinematical and dynamical characterisation of this MQG. This is the earliest quiescent system for which an IFU study is conducted (revealing an ordered rotational pattern in $V_{\ast}$, the first moment of the line-of-sight velocity distribution). Integral-field spectroscopy is necessary because it can distinguish between the types of AGN feedback that played a major role in quenching GS-9209: ejective feedback (expelling  much of the cold gas out into the CGM before it gets the chance of collapsing into stars, e.g. \citealt{hopkins2008,feruglio2010,maiolino2012agn_feedback}) vs preventative feedback (which does not allow new cold gas to be accreted by the galaxy usually due to halo heating, e.g., \citealt{bower2006,haloheating}). A possible kinematic parameter that can potentially give us insights about the type of AGN feedback in a galaxy's history is $\left(V/\sigma\right)_{R_{\rm eff}}$, the ratio between ordered stellar rotation and random motions, within the half-light radius ($R_{\rm eff}$). In the case of high redshift star-forming discs we have that $ \left(V/\sigma\right)_{R_{\rm eff}}>1$ \citep{forster_schreiber_review}. Galaxies that show a disc-like rotational kinematic pattern in $V_{\ast}$ but which have, in the same time, $ \left(V/\sigma\right)_{R_{\rm eff}}$ values lower than 1 could potentially indicate a violent, ejective AGN feedback in galaxy's history whereas $\left(V/\sigma \right)_{R_{\rm eff}} \sim 1.5-2$ are more likely to be associated with a gentle mode of AGN feedback (preventative). 

This paper is structured as follows: in \autoref{sec: data}, we discuss the general data analysis and data reduction pipeline, outlining the pre-processing that we have done on raw data before using it in our analysis. \autoref{sec: methods} discusses the computational and mathematical tools that we have used to analyse the photometry data and perform Multi-Gaussian expansion (MGE) parametrisation, as well as spectrum fitting and Voronoi binning, vital in deriving the subsequent resolved kinematics maps. In \autoref{sec: results}, we dedicate separate subsections to various important key results that highlight different properties of GS-9209. We discuss the implications for galaxy formation and evolution models and compare our results with those from independent observations in \autoref{sec: Discussion}. Our findings are summarised in \autoref{sec: Conclusions}, where we also give possible directions where further research might give precious insights. Throughout this paper, we adopt a flat universe cosmology with matter density parameter $\Omega_{\rm m} = 0.307$, Hubble Constant $\mathrm{H_{0} = 67.7 \ km \ s^{-1} \ Mpc^{-1}}$ as in \citet{planck2018}. We use the AB magnitude system.

\section{Data}
\label{sec: data}

\begin{figure}
    
    \includegraphics[width=\columnwidth]{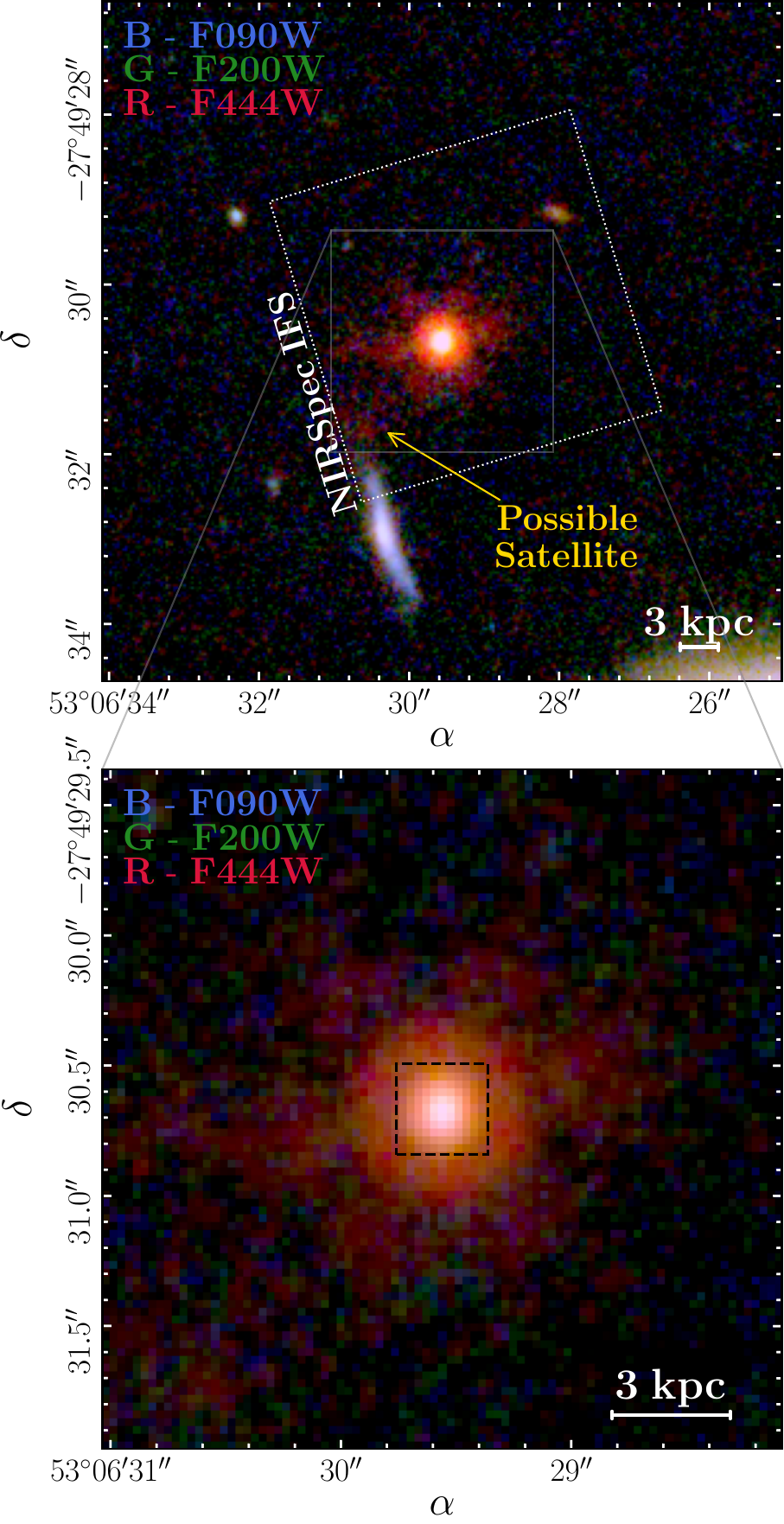}

    \caption{\textbf{Top panel:} Composite image of GS-9209 obtained by the superposition of F090W, F200W, and F444W NIRCam photometry images. The dotted line is the effective field of view of our NIRSpec
    observations. A possible satellite is indicated. \textbf{Bottom panel:} Zoomed-in portion of the same image, where we highlight the $0.35 \arcsec \times 0.35 \arcsec$ aperture used for measuring the integrated stellar velocity dispersion $\sigma_{\ast}^{\prime}$.}
    \label{fig:gs9209_RGB}
\end{figure}

The data we use were obtained as part of the {\it JWST} Cycle 2 Program ID 3659 (PI: Francesco D'Eugenio). This is a NIRSpec \citep {nirspec} IFU programme \citep{boeker22_IFs} and the observations use the high-resolution ($R \sim 2700$) disperser / filter combinations G235H / F170LP and G395H / F290LP. In this work we focus on G235H / F170LP, which was integrated for 14.7~hours, using 18 dithered observations in a 9-point medium-sized cycling pattern (repeated twice). Each dither consisted of 20 groups per integration and two integrations, using the NRSIRS2 readout mode \citep{rauscher+2017}. The medium-size dither pattern gives an effective field of view (FoV) of our particular observation of $101 \times 91$ spaxels with a resolution of $0.05 \arcsec $ per spaxel. The NIRSpec FoV is displayed in Fig.~\ref{fig:gs9209_RGB} (dotted rectangle), superimposed to a false-colour NIRCam image.
At the redshift $z \sim 4.658$ of our target, we probe a rest-frame wavelength range $\mathrm{\lambda_{rest} \in \left [\SI{2934}{\angstrom} ; \SI{5603}{\angstrom} \right ]}$ with the detector gap between $\SI{4220}{\angstrom}$ and $\SI{4310}{\angstrom}$ (rest-frame wavelengths; see \autoref{fig:spectrum_fitted}).

\subsection{Data reduction and background subtraction}
\label{sec:background}
We used the publicly available \textsc{jwst} data reduction pipeline v1.12.5, with
calibration files from the context file 1193. The standard pipeline was complemented by customized
methods for snowball flagging, pink-noise correction and outlier detection, as described in \citet{2023A&A...679A..89P}.

\subsubsection{Source Detection}
\label{sec:bkg_source_detection}
We used the package \textsc{sep}\footnote{Available at \url{https://sep.readthedocs.io/en/v1.1.x/}} (Source Extraction and Photometry Python library). This first normalises the original image and subtracts a crude estimate for the background (\textsc{sep.Background}) and then extracts the sources based on a lower  detection threshold value of 3 for the signal-to-noise ratio (S/N). The output of this procedure is a binary mask that can be used across multiple wavelengths from the data-cube. We chose to make a combined mask that reunites the individual masks produced at the wavelengths corresponding to the NIRCam F200W filter (because we will do further photometric analyses in this bandpass) and to the [O~\textsc{iii}]$\lambda5007$ emission. 

In our case, since we have an extended source, we use the function \textsc{find\_galaxy} from the \textsc{MgeFit} package\footnote{Version 6.0.4, available at \url{https://pypi.org/project/mgefit/}} (\citealt{cappellari2002mgefit}) which provides us with approximate values for the galaxy's centre position, ellipticity and major axis. We are only interested in the major axis value because we select this as the size of the mask that we are using ($7 \ \rm spaxels = 0.35 \arcsec$). This is in agreement with the value where the light from the radial profiles of the galaxy 
becomes negligible. 

\subsubsection{Background Subtraction}
\label{sec:proper_bkg_subtraction}

For the background subtraction, we use our own robust pipeline. First, the algorithm does a slice-by-slice background subtraction (these are wavelength slices) for each spaxel, using the function \textsc{Background2D} from \textsc{photutils}\footnote{We used version 2.0.1, but the most recent one is available at \url{https://photutils.readthedocs.io/en/stable/}}\citep{photutils_citation}. Together with the areas of the galaxy itself (identified using the method described in Section \ref{sec:bkg_source_detection}), the spaxels with non-valid entries (NaN) are also masked for a given wavelength slice. We use \textsc{MedianBackground} from \textsc{photutils} as a background estimator, together with a $3\sigma$ clipper (\textsc{SigmaClip} from \textsc{astropy}; \citealt{astropy_collab}) that improves the robustness of the background level calculations and reduces the possibility of contamination from bright spaxels. There are also some wavelength slices for which this background subtraction procedure fails (often because of either excessive masking or due to issues with the slice itself e.g. being within the wavelength range of the detector gap). In these cases, our pipeline handles such slices by doing an interpolation based on neighbouring ``good'' slices. 

\subsubsection{Outliers Removal}
\label{sec:outliers}

We use the outlier-detection algorithm of \citet{deugenio+2024}, but even after this step, some low-level artefacts remain. To remove them, we first mask any invalid values, then apply a median filter to smooth the spectrum, and clip any $>3\sigma$ outliers. The resulting ``double-cleaned'' spectrum is shown in \autoref{fig:spectrum_fitted}, where we extracted the spectrum from a $0.35 \arcsec \times 0.35 \arcsec$ square aperture centred on the brightest spaxel of the cube median image. This image was obtained by taking the median flux value at each spaxel across all valid wavelength slices and spans approximately $5 \rm \ R_{\rm eff}$. The chosen aperture (illustrated in \autoref{fig:gs9209_RGB}) is approximately the largest extent over which we can probe spatially resolved kinematics. 

\section{Methods}
\label{sec: methods}

\subsection{Photometry Analysis and Morphology Modelling}

For kinematical calculations, we need an accurate model of the galaxy's surface brightness profile. Our approach requires Multi-Gaussian Expansion of the light profile
\citep[\textit{MGE};][]{emsellem1994mgefit,cappellari2002mgefit}. Since GS-9209 is highly compact \citep{CarnallQuenching}, we follow the methods of \citet{vanhoudt+2021}; we first use a S\'ersic model, taking advantage of its small number of free parameters (Section~\ref{sec:photometry_pysersic}), and then
re-parametrize this best-fit model using \textit{MGE} (Section~\ref{sec:mgefit}).

\begin{figure*}
    \centering
        \includegraphics[width=\linewidth]{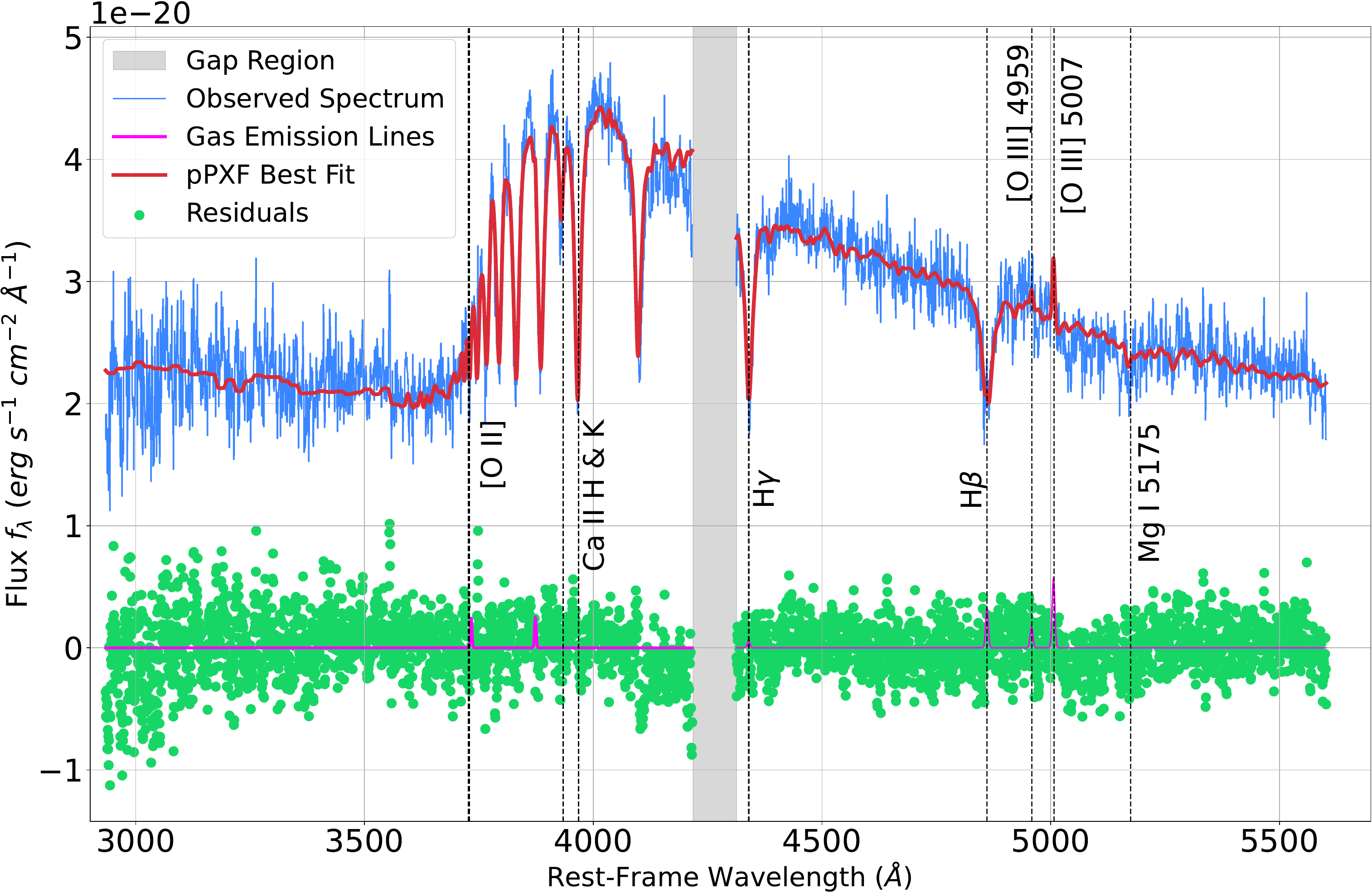}
        
        \caption{Spectrum of the galaxy GS-9209 in blue (within a $0.35 \arcsec \times 0.35 \arcsec$ square aperture) smoothed and cleaned for outliers and non-valid values. It shows a strong Balmer break of $\approx$ 2 - 2.25, indicating that old stellar populations (spectral class A and later) dominate. Some particular absorption and emission lines are highlighted. We also show the spectrum fitted with the \textsc{ppxf} algorithm (with the red colour) and the (weak) emission lines from the gas templates in magenta. Residuals are shown in green.}
        \label{fig:spectrum_fitted}
        \end{figure*}

\subsubsection{Photometry Modelling with \textsc{PySersic}}
\label{sec:photometry_pysersic}

\begin{figure*}
    \centering
    \includegraphics[width=\linewidth]{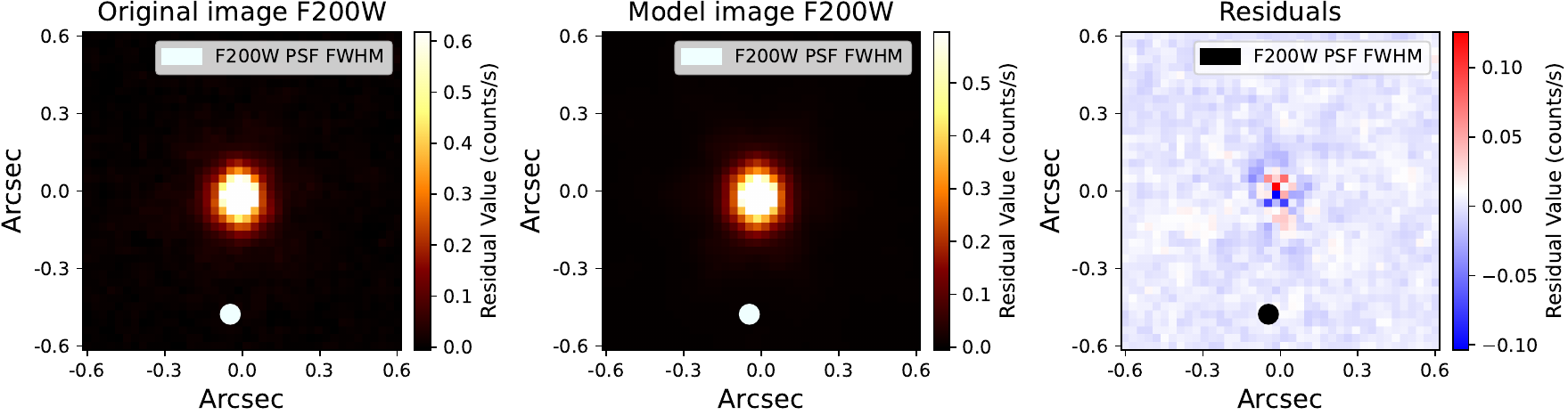}
    \caption{The original image in the NIRCam F200W band (a $40 \times 40$ pixels zoom-in around the brightest pixel; the NIRCam photometry pixel scale is $0.03 \arcsec$ per pixel) is shown in the left panel. Results of the \textsc{PySersic} fitting procedure of the original image are displayed in the middle panel. The panel on the right hand side illustrates the residuals from the fit i.e. \textit{$\rm res = \rm original - \rm model$}. On each of these images, we display the FWHM size ($0.065 \arcsec$) of the PSF of the NIRCam instrument in the F200W band.}

    \label{fig:pysersic200}
\end{figure*}

\begin{table}
\centering
\renewcommand{\arraystretch}{1.35}
\setlength{\tabcolsep}{14pt}      
\captionsetup{justification=centering}
\caption{Best fit parameters for the \textsc{PySersic} F200W Image - SVI flow method. Here $\theta$ is the astronomical PA measured by \textsc{PySersic} from North (the Y axis) counter-clockwise.}

\begin{tabular}{ c  c }

\hline
\textbf{Parameter} & \textbf{Value} \\ 
\hline
observed ellipticity & $0.251 \pm 0.015$ \\ 

S\'ersic index & $2.97 \pm 0.14$ \\ 

$R_{\rm eff} $ 
{(arcsec)} & $0.0344 \pm 0.0006$ \\ 

$\theta$ 
{(deg)} & $173.9 \pm 14.7$ \\ 

\hline
\end{tabular}

\label{tab:best_fit_parameters_pysersic_svi}
\end{table}

To determine our fiducial S\'ersic profile, we used the \textsc{PySersic}\footnote{version 0.1.5; \url{https://github.com/pysersic/pysersic}} Bayesian fitting tool for galaxy photometry \citep{Pasha2023pysersic}. We fit the galaxy image in the NIRCam F200W band and we use the empirical NIRCam ePSF from \citet{ji_core_progenitors} for this filter. The imaging data was obtained as part of JADES \citep[\textit{JWST} Advanced Deep Extragalactic Survey;][]{eisenstein2023overview}, processed as described in \citet{rieke2023jades,eisenstein2023jadesorigins,deugenio2025jadesDR3}. We show our results from \textsc{PySersic} fitting in \autoref{fig:pysersic200}; we will construct all our subsequent analysis based on the median values of the posterior distributions of the parameters of the model profile (given in Table \ref{tab:best_fit_parameters_pysersic_svi}).

In our analysis, we fit the background-subtracted NIRCam galaxy image to a S\'ersic profile \citep{sersic}. The noise maps required as inputs for \textsc{PySersic} are taken from the background-removed photometry data. In order to evaluate the quality of the models, we have chosen the Student's $t$ loss function which is to be minimised by the fitting algorithm. For estimating the posterior distribution, we compare both of the methods currently implemented into \textsc{PySersic}: Laplace method and SVI (stochastic variational inference) flow. The first method initially finds the Maximum A Posteriori (MAP) values of the parameters and assumes Gaussian posterior distributions around the MAP, whereas in the second (which we choose as fiducial), a normalised flow is trained to estimate the posterior distributions. We find that none of the results of our analysis depends on the method chosen; these results are consistent with each other, within their associated uncertainties (given in Tables \ref{tab:best_fit_parameters_pysersic_svi} in the main text and \ref{tab:best_fit_parameters_pysersic_laplace} in Appendix \ref{sec: appendix 1}). The calculated effective radius is in agreement with the value obtained by \citet{CarnallQuenching}, but our S\'ersic index $2.97\pm0.14$ is somewhat larger than the literature value ($2.3\pm0.3$). 
For the next step, we used the median values of the posterior distributions from the \textsc{PySersic} fit -- SVI method -- in order to obtain the model image, de-convolved from the empirical PSF of the original F200W image. We also oversample by a factor of 10 with respect to the original image.

\subsubsection{Multi-Gaussian Expansion (MGE) Fitting of Photometry}
\label{sec:mgefit}

We perform a \textit{MGE} fitting of the PSF de-convolved model image from \textsc{PySersic}. We used the \textsc{mge\_fit\_1d} procedure within the \textsc{MgeFit} package \citep{cappellari2002mgefit}. This procedure gives an approximation of the one-dimensional radial surface brightness profile (with a \textit{MGE} re-parametrisation) of the model image. Because the latter is, by construction, a purely S\'{e}rsic model, the surface brightness profile is given by \citep{ciotti1999sersic}: 

\begin{equation}
    I \left( R\right) = I_{0} \cdot \exp \left[-b_{n} \cdot \left (R/R_{\rm eff} \right)^{1/n} \right].
\end{equation}
 
\noindent In our case, since our procedure fits a normalised image ($L_{\rm tot}=1$), we will need to convert to real physical units after doing the fit. We can obtain $I_{0}$ using the equation 4 from \citet{ciotti1999sersic}. In that equation, $b_{n} =5.61$ for n=2.97 (equation 18 from the same paper) and $\Gamma \left(2n \right) \approx 108.36$ for n=2.97 (equation 7). With $R_{\rm eff} = 11.45$ pixels on the oversampled image, we get $I_{0} = 0.106$ normalised arbitrary units per $\rm pixel^{2}$.

Each Gaussian \textit{j} has three parameters:

\begin{itemize}
\item The peak flux $F_{j}$ given by the \textsc{MgeFit}. This is measured in arbitrary units (relative to the normalised total luminosity) per pixel.
\item The dispersion of the 2-dimensional Gaussian along the major axis, $\sigma_{j}$ in pixels.
\item The observed axial ratio of the 2-dimensional Gaussian, $q_{j}^{\prime}$ following the notation of \citet{cappellari2008jam}. In the \textsc{mge\_fit\_1d} fitting procedure, this is not given as an output of the fit. We assume that all of the 2-dimensional Gaussians have a fixed observed axial ratio $q_{j}^{\prime} = 0.749$, the value given by the \textsc{PySersic} photometry fitting tool.
\end{itemize}

GS-9209 has an apparent F200W band AB magnitude of $\rm m_{\rm F200W}=23.9 \ \rm mag$. At the source redshift z = 4.658, this corresponds to an absolute magnitude $\rm M_{\rm abs}= -22.38$ at a rest frame wavelength $\lambda_{\rm rest} \approx \SI{3510}{\angstrom}$. The pivot wavelength of the F200W bandpass is $\SI{19875}{\angstrom}$ and the absolute magnitude of the Sun in the \textit{SDSS\_u} bandpass (with a pivot wavelength of $\SI{3556}{\angstrom}$, close to the rest frame wavelength that we are probing in the case of GS-9209) is $\rm M_{u, ~\odot} = 6.39$ \citep{sun_abs_mag}. This means that the absolute luminosity of GS-9209 (at rest-frame wavelength of $\sim \SI{3500}{\angstrom}$) is $\mathrm{L_{\rm gal,tot} = 3.4 \cdot 10^{11} L_{\odot}}$. 
We then need to convert the integrated flux values that we have for our Gaussians, $F_{j}$ to surface brightness values, that we will use in our Jeans Anisotropic Modelling (\textit{JAM}) algorithm. This is done using the following equation \citep{cappellari2002mgefit, cappellari2008jam}:
{\large
\begin{equation}
    I_{j} = \frac{F_{j}}{\sigma_{j}\sqrt{2\pi}} .
\end{equation}
}


\subsection{Spectrum Fitting with \textsc{ppxf}}
\label{sec:pPXF_fitting}

To measure the stellar kinematics, we use the \textsc{ppxf}\footnote{We used version 9.4.1; the software is available from \url{https://pypi.org/project/ppxf/}} software, an efficient full spectral fitting algorithm \citep{cappellari2004ppxf,cappellari2017ppxf,cappellari2023ppxf}. \textsc{ppxf} is able to simultaneously fit both the stellar continuum and gas emission lines so one can extract the kinematics of multiple components \citep[e.g.,][]{Westfall2019,Belfiore2019}. The algorithm constructs a model of the galaxy as a linear combination of individual spectral templates, convolved with the line-of-sight velocity distribution (LOSVD). 

In our work, we used templates built using \href{https://github.com/cconroy20/fsps}{FSPS v 3.2} \citep{fsps1, fsps2}, with the \textit{MILES} stellar library for the optical region \citep{sanchezblazquez2006, falcon2011} and with the \textit{MIST} isochrones \citep{choi2016mist}. We adopted a \citet{salpeter1955} initial mass function (IMF), but this choice has no impact on the derived stellar kinematics. This is because for the stellar-population age of the target ($\sim 500$ Myr; \citealt{CarnallQuenching}) and spectral range of our data we do not track a wide enough variety of IMF-sensitive features. In order to determine the IMF characterising GS-9209, we need to use the $\rm M_{\ast}/\rm L$ ratio result from the \textit{JAM} dynamical modelling (Section \ref{sec:dynamical_modelling}). 

The stellar spectral templates have an instrumental resolution with constant full-width at half maximum (FWHM) $\Delta \lambda_{\mathrm{temp}} \approx \SI {2.5}{\angstrom}$ \citep{falcon2011}. For comparison, the average rest-frame instrumental resolution that we have for the galaxy spectra observations is $\Delta \lambda_{\rm instr} \sim \SI{1.5}{\angstrom}= \langle \lambda \rangle / \left \langle R \right \rangle $ with $\mathrm{\langle \lambda \rangle = \sqrt{\lambda_{\mathrm{min,rest}} \cdot \lambda_{\mathrm{max,rest}}}}$ and $\left \langle R \right \rangle \sim 2700$. 
Since the resolution of the data is better than the resolution of the templates, we need to correct the derived stellar dispersion using the following equation (adapted from equation 5 from \citealt{cappellari2017ppxf}): 

\begin{equation}
\mathrm{\sigma_\ast^{\prime 2} = \sigma_{\ast,\rm ppxf}^{2} - \sigma_{\rm instr}^{2} + \sigma_{\rm templ}^{2}},
\label{eq:sigmas}
\end{equation}

\noindent where $\sigma_{\rm \ast,ppxf}$, the velocity dispersion indicated by the \textsc{ppxf} algorithm, is not the same as the intrinsic (but not de-projected yet) velocity dispersion, $\sigma_{\ast}^{\prime}$. We need to correct for the Gaussian dispersions of the line spread functions of the instrument observing the galaxy ($\sigma_{\rm instr}$) and of the stellar templates that we used in our analysis ($\sigma_{\rm templ}$). In order to convert from $\Delta \lambda_{\mathrm{templ}}$ to $\sigma_{\mathrm{templ}}$ or from $\Delta \lambda_{\mathrm{instr}}$ to $\sigma_{\mathrm{instr}}$ we use the general formula (valid for both cases):

\begin{equation}
\mathrm{\sigma_{\rm j} = \frac{c \cdot \Delta \lambda_{\rm j} \ / \lambda_{\rm med}}{2.355}},
\end{equation}

\noindent where ``j'' = either ``templ'' or ``instr'' and $\lambda_{\mathrm{med}}$ is the median wavelength of our range. To confirm the accuracy of our methods, we repeated our analysis by using stellar templates from a higher-resolution library\footnote{These templates are available from the authors C. Conroy upon reasonable request.}($R\sim3200$) which uses the \textit{C3K} model atmospheres \citep{conroy2019c3k} instead of \textit{MILES}. The results did not differ by more than a few percent compared to our original method (e.g. for the spectrum in \autoref{fig:spectrum_fitted} using \textit{MILES} stellar spectra, we obtained $\sigma_{\ast}^{\prime} =239 \pm 34$  \kms compared to $234 \pm 37$ \kms when using \textit{C3K}). This is in good agreement with the value obtained by \citet[][using medium-resolution spectroscopy]{CarnallQuenching} for the stellar dispersion $\sigma_{\ast}^{\prime} = 247 \pm 16$ \kms. 

\subsection{Voronoi Bins Analysis}
\label{sec:vorbin_analysis}



The purpose of the Voronoi binning algorithm is to perform a 2-dimensional re-binning of our spectral data to achieve a minimum S/N threshold. We create S/N maps for all the spaxels using the median-filtered signal across the wavelength slices for which $\mathrm{\lambda_{rest} < \SI{3500}{\angstrom}}$. As can be seen from \autoref{fig:spectrum_fitted} this represents the flattest region of the spectrum, so the computed S/N is not affected too much by the presence of absorption/emission lines or by the slope of the continuum. We mask all the noise dominated spaxels with $\mathrm{S/N \ < 1} $. We used the \textsc{VorBin}\footnote{We used version 3.1.5, available from \url{https://pypi.org/project/vorbin/}} package \citep{cappellari2003vorbin}. We set a threshold S/N target $>3$ for our Voronoi bins. 

For each Voronoi bin, we obtain the spectrum as the sum of the spectra of the pixels contained within that bin. The spectrum for each Voronoi bin has been further processed as follows:

\begin{itemize}
    \item In the case of each spectrum, the non-valid values and the $> 3 \sigma$ outliers were masked. 
    This creates the ``cleaned'' Voronoi bin spectrum, which is then smoothed using a median filter. The smoothed spectrum is not brought forward into our analysis but it is used only to identify and mask bad spectral pixels. 
    \item We log-rebin each ``cleaned'' spectrum to a velocity scale $c \times \ln \left(\lambda_{\rm max}/\lambda_{\rm min} \right)/N_{\lambda} \approx$ 50 \kms (here $N_{\lambda}$ represents the number of wavelength slices) and run an initial \textsc{ppxf} fit. We impose constraints on the ages and metallicities ranges of the templates to be used: age $<$ 1.3 Gyr (the age of the Universe at $z \sim 4.66$) and $-1 \leq \mathrm{log} \left ( Z/ Z_{\odot} \right) \leq 0.5$. The purpose of allowing high-metallicity templates is to capture the effect of $\alpha$-elements enrichment
    \citep[e.g.,][]{beverage+2021,beverage+2023,carnall2023excels}. The goal of this initial \textsc{ppxf} run on the cleaned spectra is to reduce the numbers of stellar templates that we use. We determine the stellar templates (in this first run of the \textsc{ppxf} fit) for which the light weights are positive. These stellar templates will be used in the second fit (the gas templates remain unchanged).
    \item We then run a second \textsc{ppxf} fit on the ``cleaned'' Voronoi bin spectra and the results of this fit are used for further analysis, specifically the stellar kinematic values $V_{\ast, \rm ppxf}$, the stellar rotational velocity, and $\sigma_{\ast, \rm ppxf}$, the stellar velocity dispersion returned by \textsc{ppxf}, together with the uncertainties in each of these parameters. We use equation \ref{eq:sigmas} in order to take into account the difference between the stellar velocity dispersion returned by \textsc{ppxf} and the value corrected for instrument and templates line widths. 
\end{itemize}
 
Using the intrinsic velocity dispersion (obtained after applying the correction from Equation \ref{eq:sigmas}) we build our $V_{\rm rms}$ map that is required for our subsequent \textit{JAM} Dynamical Modelling (\autoref{sec:dynamical_modelling}). Here $V_{\rm rms} = \sqrt{V_{\ast}^{2} +\sigma_{\rm \ast}^{2}}$.

\section{Results}
\label{sec: results}
\subsection{Spatially Resolved Kinematics Maps of GS-9209}
\label{sec:results_kinematics}

\begin{figure*}
    \centering

    \centering
    \includegraphics[width=\linewidth,height=8cm]{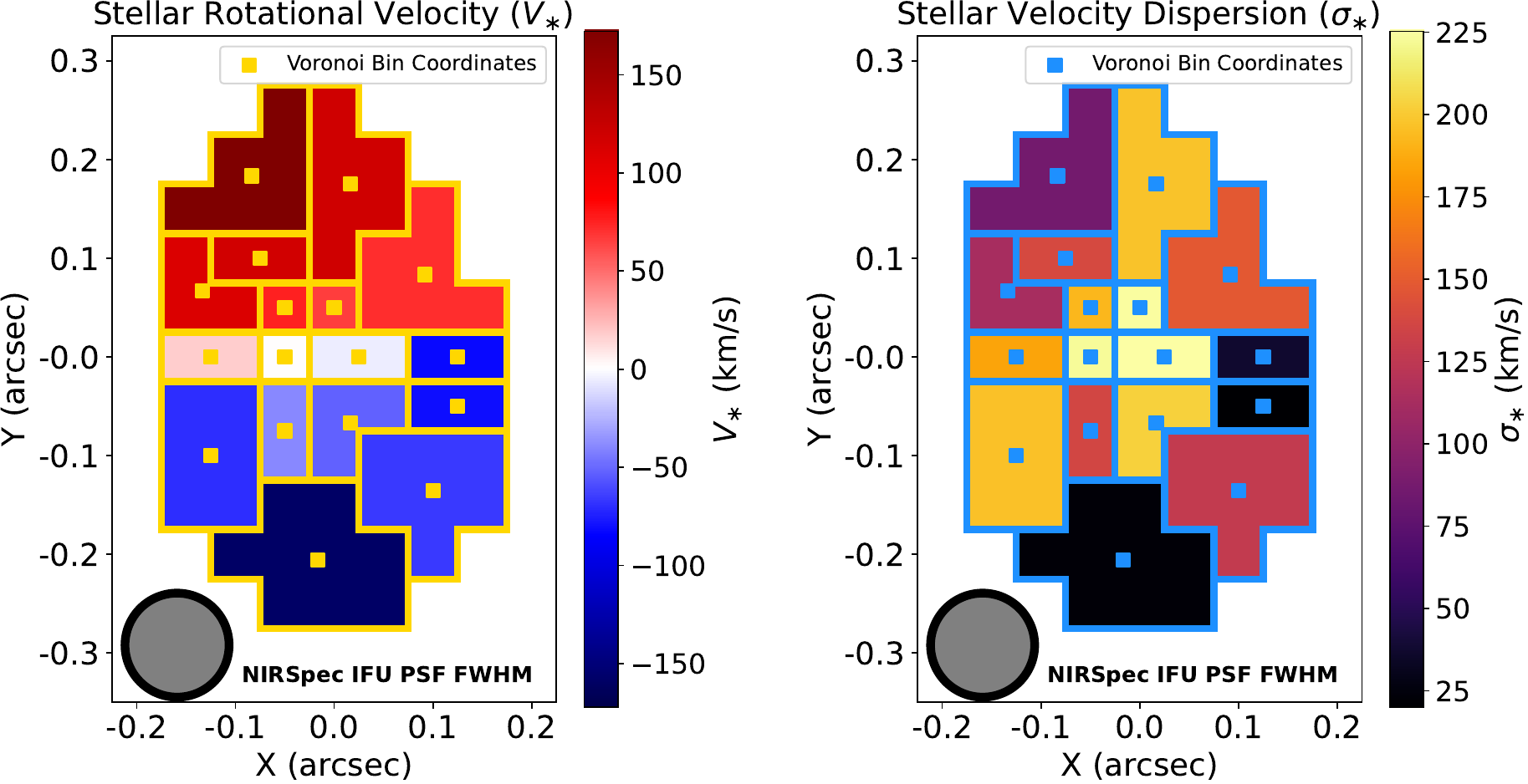}
    \caption{Maps of the observed (i.e. not de-projected) first (left panel) and second (right panel) moments of the stellar kinematics for the massive quiescent galaxy GS-9209 i.e. the stellar line-of-sight velocity ($V_{\ast}$) and the stellar velocity dispersion ($\sigma_{\ast}$). The left panel indicates a rotational pattern of the motions of the stars in this galaxy. The right panel shows a dispersion dominated central region. The spatially resolved kinematics maps extend to about $4 \ \rm spaxels=0.2\arcsec$, which means approximately $6 \ R_{\rm eff}\approx 1.3 \ \rm kpc$. We furthermore show the outlines separating the Voronoi bins from each other and the positions of the centres of each Voronoi bin. The spaxels contained within each Voronoi bin are all colour coded according to either $V_{\ast}$ (left panel) or $\sigma_{\ast}$ (right panel). The plots also display the PSF FWHM of the NIRSpec IFU instrument at $\lambda_{\rm obs} =2 \ \mu m$.}

    \label{fig:kinematics_results_new}
    

\end{figure*}

The results of our stellar kinematics analysis are shown in \autoref{fig:kinematics_results_new}. The image on the left side shows that the stars in this galaxy have substantial angular momentum around the galaxy centre. 
The image on the right side shows a decrease of the stellar velocity dispersion from the galaxy centre to its outskirts. With our kinematics data alone, it is difficult to disentangle between beam smearing effects and the plausible explanation of this structure being related to the presence of a dispersion dominated central bulge. This information can be revealed if we combine spatially resolved kinematics data and resolved stellar population ages maps. 
The key point of this analysis is that the disc-like kinematics is still preserved after the quenching took place. Hence it turns out that galaxies may first stop forming stars (and quench inside-out) but preserve their disc-like kinematics for another time interval before other dynamical processes may come into play and transform them into the slowly rotating spheroids and early-type galaxies (ETGs) that we observe at $z \sim 0$.

\subsection{JAM Dynamical Modelling}
\label{sec:dynamical_modelling}

\begin{figure*}
    \centering
    \includegraphics[width=\linewidth]{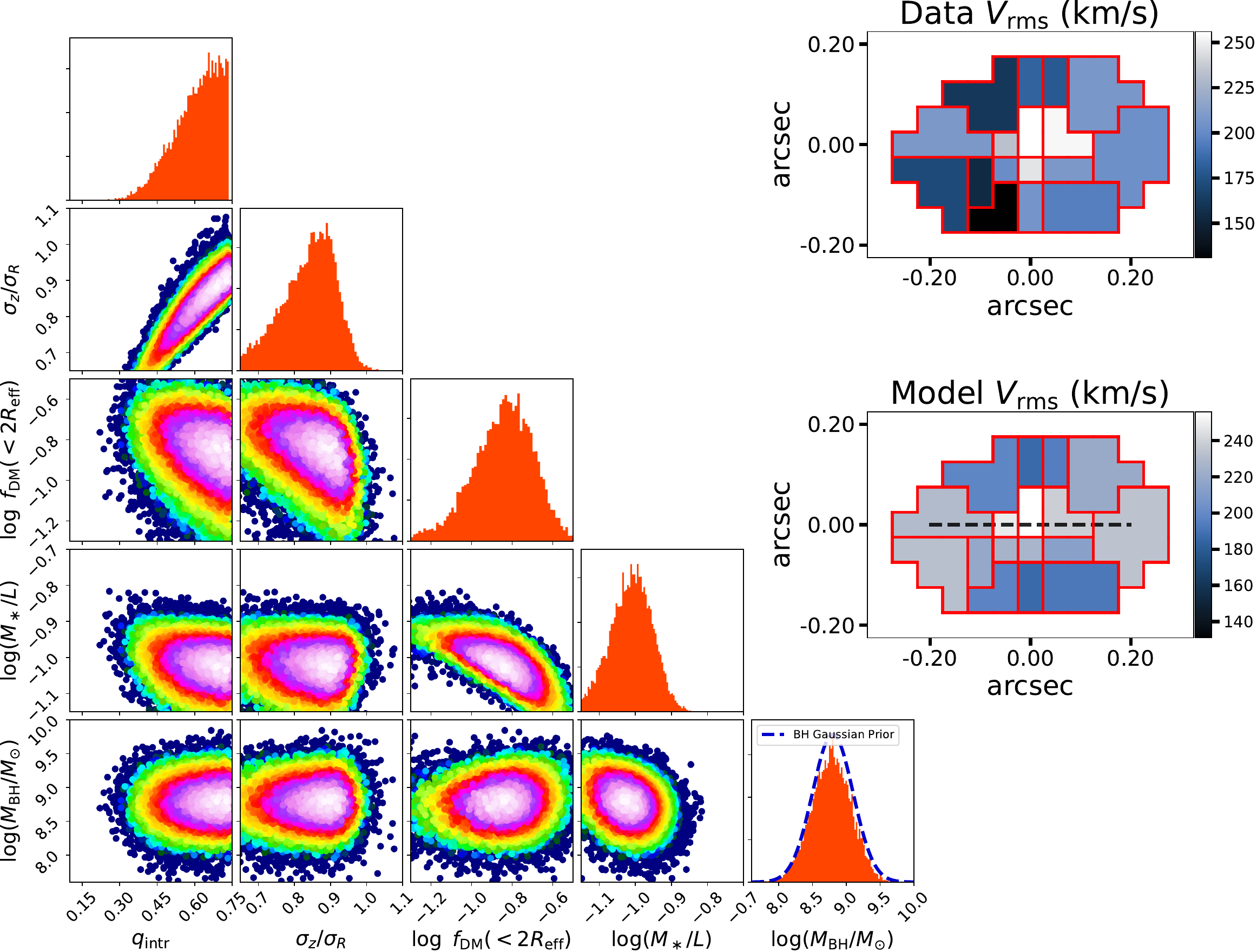}
    \caption{Outputs of the fiducial dynamical model. In this model, we consider a Gaussian prior on the mass of the central BH (as shown on the bottom right corner plot panel; see the details in \autoref{sec:dynamical_modelling}) and we furthermore assume a classic NFW density profile of the dark matter halo. We obtain a Dark Matter fraction of about $14.5^{+6.0}_{-4.2}$\% within two effective radii. 
    In this plot, we denote $\log \left(M_{\ast}/L\right) \equiv \log \left[ \left(M_{\ast}/L\right)/\left(M_{\odot}/L_{\odot}\right) \right]$. The fiducial dynamical model has $\chi_{r}^{2}=1.4$. The central black dashed line corresponds to the major axis of the galaxy which we will use in the calculation of $\left(V/\sigma\right)_{\ast}$ in Section \ref{sec:V_sigma}.}

    \label{fig:adamet_fiducial}
\end{figure*}

\begin{table}
\centering
\renewcommand{\arraystretch}{1.35} 
\setlength{\tabcolsep}{11pt}    
\caption{Marginalized posterior probabilities on the free parameters of our fiducial dynamical model (which assumes a NFW Dark Matter density profile and a Gaussian prior on $\log M_{\rm BH}$; see the main text). The values reported are the median and 16\textsuperscript{th}--84\textsuperscript{th} percentile probability range.}

\begin{tabular}{ c c c}
\hline
Parameter & Value & Model \\ 
\hline 
$q_{\rm intr}$ & $0.62 \pm 0.10$ & \textit{Fiducial} \\
$\sigma_{z} / \sigma_{R}$ & $0.85 \pm 0.08$  & \textit{Fiducial} \\ 
$\log f_{\rm DM} \left(<2R_{\rm eff} \right)$ & $-0.84 \pm 0.15$ & \textit{Fiducial} \\
$\log \left[ \left(M_{\ast}/L \right)/\left(M_{\odot}/L_{\odot}\right) \right]$ & $-1.01 \pm 0.06$ & \textit{Fiducial} \\
$\log M_{\rm BH}/M_{\odot}$ & imposed prior & \textit{Fiducial} \\

\hline

\end{tabular}
\label{tab:table_fiducial_alternative}
\end{table}

Jeans Anisotropic Modelling is a robust method that is suitable for a plethora of galaxies dynamical modelling tasks including estimating the galaxies dark matter content \citep{van_de_ven2010,barnabe2012,ngc1277}, 
computing central supermassive black holes masses \citep{drehmer+2015,feldmeier+2017,ngc1277} or determining dynamical masses to high accuracy \citep{van_der_Marel_2007,michele2009_dynamicals,vanhoudt+2021,dynpop-1,cappellari2023ppxf}. The better performance and accuracy of \textit{JAM} has been tested and acknowledged against other techniques such as the more generalised Schwarzschild dynamical models \citep{schwarzschild1979}. Such tests were done using either simulated data \citep{jin+2019} or real data \citep{leung+2018}. These tests revealed the ability of this method to accurately produce a robust dynamical model for a wide variety of input kinematics data. In our study, we use the \textsc{JamPy} package\footnote{version 8.0.0, available at \url{https://pypi.org/project/jampy}}.

To calculate the posterior probability distributions on the model parameters, we use the Adaptive Metropolis algorithm \citep{metropolis1953} as implemented within the \textsc{adamet} package\footnote{version 2.0.9 available from \url{https://pypi.org/project/adamet/}}  \citep{atlas15}.  We repeated our analysis using the Markov-Chain Monte-Carlo sampling alogirthm \textsc{emcee} \citep{emcee_paper2010,emcee_paper2012} and we concluded the results obtained were statistically consistent but that \textsc{adamet} algorithm had a better time efficiency by a factor of $\sim 2$. 

Our \textit{JAM} dynamical models rely on the assumption that our galaxy is axisymmetric and furthermore that there is no spatial variation of the $M_{\ast}/L$ ratio or of the radial anisotropy parameter \citep{atlas12}. This assumption is motivated by the findings of \citet{suess+2021} who report that massive PSB galaxies at $2<z<2.5$ typically have flat colour gradients. The outputs from the \textit{MGE} fitting of the F200W photometry (surface brightness, dispersions along the major axis and observed axial ratios of the Gaussians) are given as inputs for the light map of \textit{JAM}. The choice of F200W band to construct these flux maps is justified because in this way, we capture the stellar light at rest-frame wavelengths near the Balmer break ($\lambda_{\rm rest}\sim\SI{3500}{\angstrom}-\SI{4000}{\angstrom}$) primarily coming from type A-stars dominating the spectrum of GS-9209. Furthermore, at these wavelengths, we avoid possible contamination from faint AGN that might be present at lower rest-frame wavelengths. 

The coordinates of the Voronoi bins that we pass onto \textit{JAM} were rotated such that the major axis of the galaxy is parallel to the horizontal axis of the new coordinate system. This means that we need to rotate our $x_{\rm bin}$ and $y_{\rm bin}$ counter-clockwise by an angle of $-98^{\circ}$ (the best-fit position angle retrieved by the \textsc{pafit} algorithm\footnote{version 2.0.8 available from \url{https://pypi.org/project/pafit/}}). This value agrees to within $2^{\circ}$ with the negative of the position angle measured clock-wise from the image X-axis. i.e. $270^{\circ} - \theta$ where $\theta$ is the astronomical PA measured by \textsc{PySersic} from North (the Y axis) counter-clockwise, suggesting a very good alignment of the kinematic axis to the morphological major axis. Our \textit{JAM} dynamical modelling algorithm further requires the size of the PSF as an additional input. We assume a 2-dimensional Gaussian PSF of circular shape i.e. with the same FWHM in both the direction of the NIRSpec/IFU slicers and perpendicular to them. The NIRSpec PSF size that we give as an input to \textit{JAM} is the geometrical average between the values given by the equations 3 and 4 from \citet{deugenio+2024} adopting a typical wavelength of $\lambda \sim 2 \ \mu m$ (because our photometry analysis was based on NIRCam F200W data). In our case, this average value is $\rm FWHM_{\rm PSF} = 0.104 \arcsec$. 

All the dynamical models that we build assume a cylindrically aligned framework of the velocity ellipsoid \citep{cappellari2008jam}. Our fiducial dynamical model assumes a Navarro-Frenk-White (NFW; \citealt{nfw1996}) density profile for the dark matter halo and a Gaussian probability prior on the logarithm of the central BH mass (with the mean being  $\langle \log \rm M_{\rm BH}/M_{\odot} \rangle = 8.8$ and a standard deviation of 0.3). The choice for the mean and the standard deviation of this Gaussian prior is motivated by the results from the broad line component of $H\alpha$ fit obtained by \citet{CarnallQuenching}. The reason for imposing a prior on the central BH mass is because our spatial resolution of $0.05 \arcsec \ \rm spaxel^{-1}$ is about 6.5 times larger than the radius of the black hole's sphere of influence (assuming the mass obtained by \citealt{CarnallQuenching}). This is discussed in Appendix \ref{sec:appendix_BHMFL} and illustrated in \autoref{fig:adamet_BH_MFL} where we present our \textit{Alternative-1} Black-Hole-Mass-Follows-Light (BH+MFL) model used to infer the central BH mass, which was allowed to be fit freely by the \textit{JAM} algorithm. The other parameters of \textit{Alternative-1} the dynamical model (reported in \autoref{tab:table_just_alternative1}) 
agree within $1\sigma$ uncertainties with the values obtained from our \textit{Fiducial} dynamical model (the latter were reported in \autoref{tab:table_fiducial_alternative}). 

We obtain a stellar mass for this galaxy of $\mathrm{\log \left (M_{\ast} / M_{\odot} \right ) = 10.52 \pm 0.06}$ from our \textit{Fiducial} dynamical model. This value is consistent within $1\sigma$ with the value of $10.58 \pm 0.02$ estimated by \citet{CarnallQuenching} from stellar population modelling. In any case, we can see from e.g. figure 1 of \citet{Qiao45to85} that this galaxy is at least $3\sigma$ more massive than the median of the redshift range $4.5 \leq z \leq 6.5$.

In our fiducial dynamical model, the dark matter density profile is assumed to be classical NFW (with the small radius slope $\gamma = -1$) satisfying the equation \citep{nfw1996,lenses2001general}:
{\large
\begin{equation}
\mathrm{\rho_{\rm DM} \left ( r \right ) \propto \frac{1}{\left (r  /   r_{b} \right )^{-\gamma} \cdot \left (1 + r  /  r_{b} \right ) \ ^{3+\gamma}}} ,
\label{eq:nfw_profile}
\end{equation}
}

\noindent $\mathrm{r_{b}}$ is the break radius of the dark matter (DM) Halo in which GS-9209 resides. Using the stellar mass $-$ halo mass (SMHM) relation results from the figure 7 of \citet{behrooziDM}, we deduce that the halo of GS-9209 has a mass of $\mathrm{\log \left ( M_{\rm h} / M_{\odot} \right )} \approx 12.36 \pm 0.10$. 
Then we determine the virial radius of our DM Halo, $R_{\rm vir}$ from: 
{\large
\begin{equation}
\mathrm{M_{h} = \frac{4 \pi}{3} \Delta_{vir} \rho_{m} \left ( z \right )  R_{vir}^{3}} .
\label{eq:mvir_rvir}
\end{equation}
}
\noindent In this equation, we assume that the density contrast is $\Delta_{\rm vir} = 200$. Doing the calculation we obtain $R_{\rm vir} \approx 70 \pm 6$ kpc. Next we estimate the concentration parameter $\mathrm{c_{200} = R_{vir}  /  r_{b}}$ from figure 5 of \citet{klypin2011}. We obtain a value of $c_{200} = 4.7 \pm 0.2$ which gives $r_{b} = 14.9 \pm 1.4$ kpc. This is the reason for which we have chosen to fix this break radius in our calculations to 15 kpc ($2.31 \arcsec$ at this redshift) in our \textit{JAM} dynamical modelling algorithm. In a similar way to the method illustrated in Section \ref{sec:mgefit}, we fit this NFW profile to obtain the \textit{MGE} Gaussians for the Dark Matter component. Subsequently, the one-dimensional \textit{MGE} best-fit of the dark matter halo density profile and the \textit{MGE} fit of the stellar component light profile (constructed in Section \ref{sec:mgefit}) are combined. 

The dynamical model is used to determine a number of parameters that define either the morphology or the structure of this galaxy:
\begin{itemize}
    \item $\mathbf{q_{\rm intr}}$, the intrinsic axial ratio. Combined with the observed axial ratio, this will give us the inclination angle of the galaxy with respect to the line-of-sight. The latter is crucial in computing the de-projected values of various kinematic parameters (\autoref{sec:kinematics_quant}).
    \item $\boldsymbol{\sigma_{z} \ / \ \sigma_{R}}$, the radial anisotropy of the velocity ellipsoid (cylindrically aligned). This is also used in de-projection calculations.
    \item $\mathbf{M_{\ast}/L}$, the stellar mass to light ratio. 
    
\end{itemize}

Because $V_{\rm rms}$ is independent of the tangential anisotropy (see the discussion in Section 3.1.5 of \citealt{cappellari2008jam}), we need to test another dynamical model \textit{Alternative-2} in order to determine the tangential anisotropy $\sigma_{\phi}/\sigma_{\rm R}$ (this parameter is required for subsequent de-projection calculations). We fit the first order velocity moment along the line-of-sight i.e. the $V_{\ast}$ spatially resolved map from the left panel of \autoref{fig:kinematics_results_new} with the \textit{JAM} keyword input \textit{moment=`z'}. In this model, we again assume a classic NFW DM halo density profile and Gaussian priors on each of the parameters $q_{\rm intr}$, $\sigma_{\rm z}/\sigma_{\rm R}$, $\log f_{\rm DM} \left(<2R_{\rm eff} \right)$, $\log \left [\left(M_{\ast}/L\right)/\left(M_{\odot}/L_{\odot}\right) \right]$ and $\log \left (M_{\rm BH}/M_{\odot} \right)$. The means and the standard deviations of their Gaussian priors are taken as the best-fit and the uncertainty estimates produced as outputs of \textit{Fiducial} dynamical model (Table \ref{tab:table_fiducial_alternative}). We introduce the new parameter $\sigma_{\phi}/\sigma_{\rm R}$ to be fitted by the \textit{JAM} algorithm. We impose an informative Gaussian prior on this parameter with a mean of 1 and a standard deviation of 0.2 as motivated by the second panel of figure 2 from \citet{sauron10}. The outcome posterior distributions of this model are shown in \autoref{fig:tangential_anisotropy} in Appendix \ref{sec:appendix 2}. We are thus able to put a tight constraint $\sigma_{\phi}/\sigma_{\rm R} =1.01 \pm 0.05$.

\subsection{Dynamical Mass of GS-9209}
\label{sec:dynamical_mass}

We compute the total dynamical mass of this galaxy using equation 7 from \citet{vanderWel2022} (assuming that our galaxy behaves kinematically similar to a disc; this is justified by the rotation pattern that we can see in spatially resolved $V_{\ast}$ map; \autoref{fig:kinematics_results_new}):

\begin{equation}
    \mathrm{M_{dyn} = K \left ( n_{ser} \right ) K \left (q^{\prime} \right ) R_{\rm eff} \sigma_{\ast}^{\prime 2}  / G}.
\end{equation}

\noindent We use the same parametrisations as in \citet{vanderWel2022}; this parametrisation assumes that the total dynamical mass is two times larger than the mass predicted by \textit{JAM} dynamical modelling within the half light radius: $\rm M_{\rm dyn} = 2 \times \rm M_{\rm JAM} \left(< \rm R_{\rm eff} \right)$. In this equation, $K(n_{\rm ser})$ is given by eq.~(20) of \citet{Cappellari2006}. This correction factor depends on the S\'ersic index (we adopt a value $n_{\rm ser} = 2.97 \pm 0.14$ following the results of the \textsc{PySersic} photometric fitting and given in Table \ref{tab:best_fit_parameters_pysersic_svi}) whereas $K\left(q^{\prime}\right)$ depends on the observed axial ratio, for which we take $q^{\prime}=1-\epsilon^{\prime} = 0.748 \ \pm \ 0.015$ with $\epsilon^{\prime}$ also taken from the \textsc{PySersic} fitting results (\autoref{tab:best_fit_parameters_pysersic_svi}). From the \textsc{PySersic} photometry fitting, we have that $R_{\rm eff}  = 223 \pm 4 \  \rm pc$. This uncertainty is propagated from the \textsc{PySersic} fitting and is not our actual error for $R_{\rm eff}$. We estimate the latter to be around $\pm 20 \ \rm pc$, taking into account the systematic uncertainties. For the stellar velocity dispersion, we take $\sigma_{\ast}^{\prime}$ to be the non de-projected value that we obtained from the \textsc{ppxf} full spectrum fitting i.e. $\sigma_{\ast}^{\prime} = 239 \pm 34  \ \rm \kms$ (\autoref{sec:pPXF_fitting}).

Using these values, we obtain $\mathrm{\log \left (M_{\rm dyn} / M_{\odot} \right) = 10.3 \pm 0.3 }$. The quoted uncertainties are dominated by systematics such as the calibration uncertainty of the virial estimator scaling relation. Overall, this result is consistent (within the $1\sigma$ uncertainties) with the stellar masses that we have found using our \textit{Fiducial} dynamical model ($\log \left(\rm M_{\ast}/\rm M_{\odot}\right)$ = 10.52). A considerably more robust method of computing the dynamical mass was proposed in Section 5.2 of \citet{cappellari2023ppxf} and relies on the use of \textsc{jam\_axi\_sersic\_mass} procedure which has a number of important advantages such as allowing us to pass as inputs the size of the PSF 
and our custom rectangular aperture ($0.35 \arcsec \times 0.35 \arcsec$) that we used to produce the spectrum in Fig. \ref{fig:spectrum_fitted}. Furthermore, the method takes into account the effects of inclination and radial anisotropy when predicting the dynamical mass of the S\'{e}rsic model of our galaxy. With $n_{\rm Ser}=2.97$, and the values for $R_{\rm eff}$ and $\sigma_{\ast}^{\prime}$ mentioned above, we obtain a remarkably similar value: $\log \left (M_{\rm dyn, JAM}/M_{\odot}\right) = 10.32 \pm 0.14$. Our result is consistent with the fact the target has a relatively low dark matter fraction, as inferred from our fiducial dynamical model (\autoref{fig:adamet_fiducial}), with the total density of this galaxy being close to its stellar density. 

\subsection{Quantitative Analysis of the Galaxy's Kinematics}
\label{sec:kinematics_quant}

To allow for a better comparison with the observationally retrieved kinematics parameters of other galaxy samples from the literature or with the characteristics of galaxies in numerical simulations, we must compute some quantities that numerically quantify the rotational pattern observed in this galaxy and the net contribution of rotation itself to the total kinetic energy budget of the galaxy. These are: 
\begin{itemize}
    \item $\boldsymbol{  \left(V/{\sigma}\right)_{\ast} }$ the ratio of between stellar rotational velocity and velocity dispersion.
    \item $\boldsymbol{\lambda_{2  \rm R_{\rm eff}}}:$ the spin parameter of our galaxy, calculated using all the bins available within a projected distance of $\leq 2 \ \rm R_{\rm eff}$ from the galaxy's centre. The reason for choosing an aperture of $2 \ \rm R_{\rm eff}$ instead of the conventional $1 \ \rm R_{\rm eff}$ is because of the fact that we only have one Voronoi bin within the latter aperture. We adopt the effective radius and the galaxy's centre position that were determined by \textsc{PySersic}. 
    \item $\boldsymbol{\kappa_{\rm rot}} = \left(\rm K.E.\right)_{\rm rot}/\left(\rm K.E.\right)_{\rm total}$. This is the fraction of kinetic energy that goes into ordered stellar rotation (defined in the same way as in \citealt{sales2012}). 
\end{itemize}

\subsubsection{Inclination Angle; De-projection calculations}
\label{sec:deprojections}

The intrinsic (de-projected) axial ratio of the galaxy, $q_{\rm intr}$, is a free parameter that our \textit{JAM} algorithm provides a best-fit for. In the \textit{Fiducial} model (which is more robust compared to the BH-MFL one in Appendix \ref{sec:appendix_BHMFL}), \textit{JAM} predicts a value $q_{\rm intr} = 0.62 \pm 0.10$ (\autoref{tab:table_fiducial_alternative}), although the data do not provide tight constraints, since the posterior probability distribution is consistent with the upper limit set by the priors (\autoref{fig:adamet_fiducial}). In order to test the robustness of this estimate, we conduct a further dynamical model test \textit{Alternative-3}. In this model, we use the same setup as in the \textit{Fiducial} model but this time we put an additional Gaussian prior on $q_{\rm intr}$ with a mean of 0.41 and a standard deviation of 0.18. The choice for these values is motivated by the study of \citet{vanhoudt+2021} on a sample of massive galaxies from the LEGA-C survey \citep{vanderWel2016legac}. While these values apply to a sample of galaxies at $0.6<z<1$, there are currently no similarly large spectroscopic surveys of quiescent galaxies at much higher redshifts than this LEGA-C sample. As given in Table \ref{tab:best-fit-alternative3} the \textit{Alternative-3} dynamical model gives $q_{\rm intr,3} =0.58 \pm 0.09$ and it has a higher reduced $\chi_{r}^{2}$ than the \textit{Fiducial} dynamical model. Since $q_{\rm intr,3}$ is significantly closer to $q_{\rm intr}$ retrieved by the fiducial model than to the mean of the prior imposed on this parameter within the \textit{Alternative-3} dynamical model, it means that we can trust the reliability of our data to determine the true value of the intrinsic axial ratio of GS-9209. Hence we assume $q_{\rm intr} = 0.62$ in our subsequent calculations. This value agrees with the median axial ratio found by \citet{kawin+25} for a sample of 17 massive quiescent galaxies at $3<z<4.3$ (originally from \citealt{nanayakkara+24}).

The equation that relates the observed axial ratio, the intrinsic axial ratio and the inclination of our line-of-sight compared to the galaxy axis of symmetry is the equation 14 from \citet{cappellari2008jam}. Re-arranging the terms, we have:
{\large
\begin{equation}
    \mathrm{\tan i = \sqrt{\frac{1-q^{\prime 2}}{q^{\prime 2}-q_{intr}^{2}}}} .
    \label{eq:observed_real_inclination}
\end{equation}
}
\noindent We obtain an inclination angle of $i = 57.8^{\circ} \pm 9.5^{\circ}$ ($i =0^{\circ}$ would correspond to a face-on view). The inclination angle that we obtain for GS-9209 is surprisingly (although not physically impossibly) high. We should keep in mind the fact that this galaxy exhibits line emission from the AGN broad-line region \citep[BLR;][]{CarnallQuenching}. There have been several studies about BLRs, Type-1 AGN and Seyfert-1 galaxies from a geometrical perspective \citep{Zhang2002blr,decarli2011blr,kuhn2024BLR}. The general consensus is that the maximum half-opening angle of the cone of BLR emission is somewhere in the region $\theta_{o} \approx 50^{\circ} - 60^{\circ}$ which means that, in our case, the detection of broad line $H\alpha$ emission in GS-9209 implies that the inclination angle should (conservatively) be below $60^{\circ}$. Given the inclination angle value of $i \approx 58^{\circ}$ for GS-9209, it means that the line-of-sight is only barely inside the BLR emission cone. Although the distribution of the inclination angles for general BLRs is peaked at around $\left \langle i \right \rangle \approx 30^{\circ}$ if we assume thin disc shapes \citep{decarli2011blr}, there have been studies which reported an inclination angle close to the cone half opening angle and also close to about $60^{\circ}$ (e.g. \citealt{bentz2022blr}). In other cases, the inclination angles of the galactic disc and the inclination angle of the BLR are widely different but broad line emission is still detected (e.g. sources MCG +04--22--042, Mrk 1392 or Zw  $229-015$; \citealt{durong2024}). 

\subsubsection{\texorpdfstring{$V/\sigma$}{v/s}}
\label{sec:V_sigma}

\begin{figure}
    \centering
        \begin{subfigure}{0.48\textwidth}
            \centering
            \includegraphics[width=\linewidth]{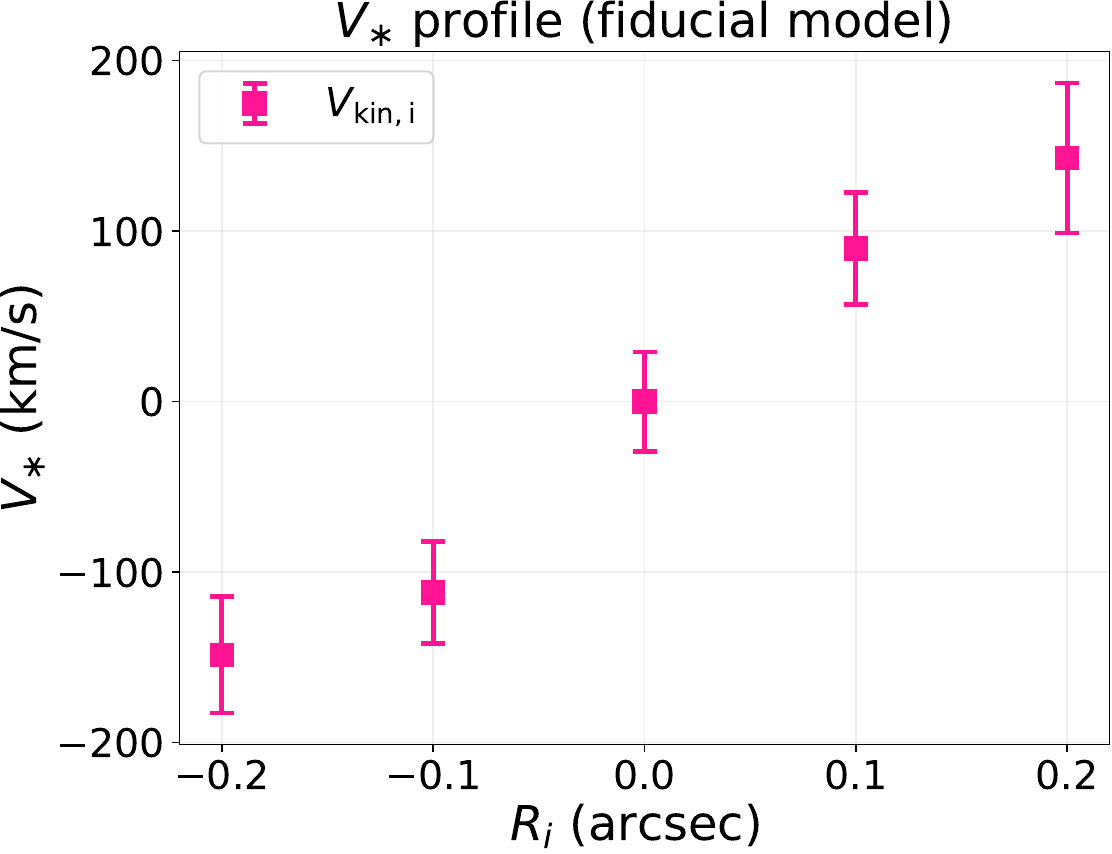}
            \vspace{-0.12cm}
        \end{subfigure}
        
        \begin{subfigure}{0.48\textwidth}
            \centering
            \includegraphics[width=\linewidth]{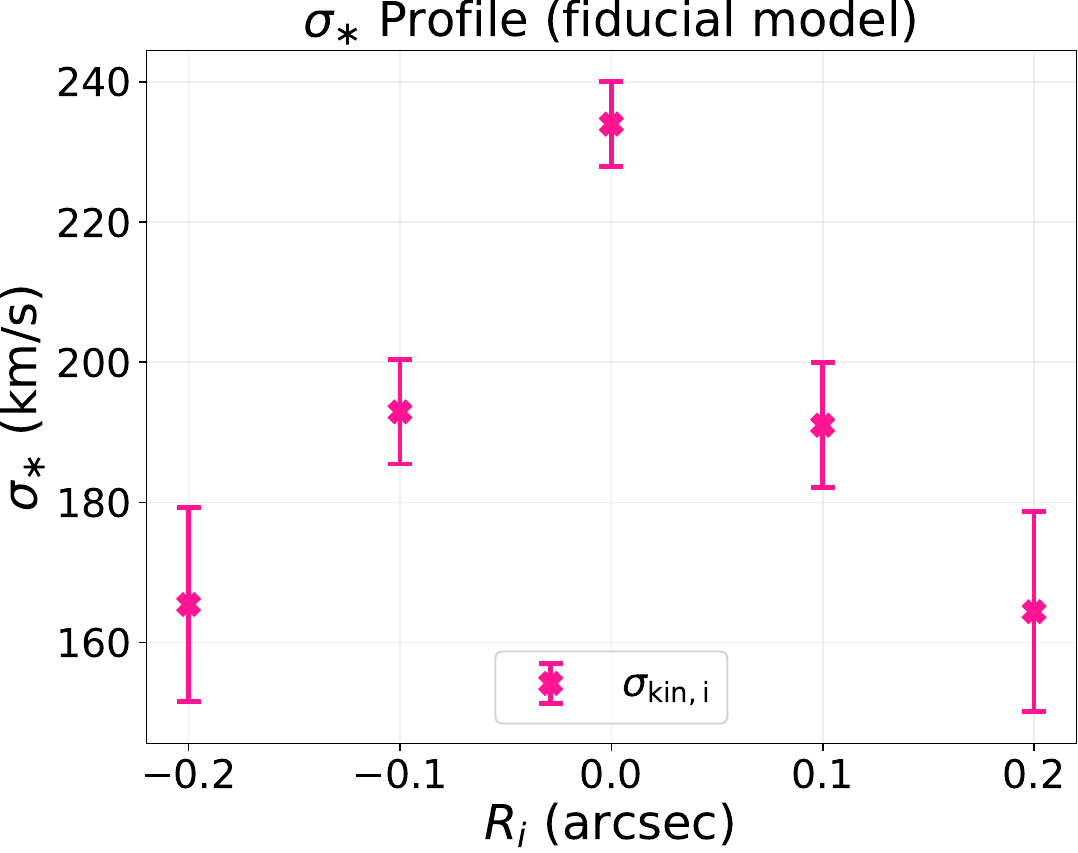}
           
        \end{subfigure}
     
        

           
           
       
    \caption{Non-deprojected stellar kinematics first and second moments ($V_{\ast}$ - top panel and $\sigma_{\ast}$ - bottom panel, together with their $1\sigma$ uncertainties) radial profiles obtained from the fiducial dynamical model.} 
    \label{fig:profiles_fiducial-1}
\end{figure}

We adopt the following method for determining $\left(V/ \sigma\right)_{\ast}$. We consider a set of points located on the major axis of the galaxy (in the rotated coordinate system in which the major axis is in the horizontal direction). Therefore, we select this set of points in the projected space such that they have $y_{\rm rot, P} = 0$ and they are equally spaced between $x_{\rm rot, P}$ = -4 spaxels and $x_{\rm rot, P}$ = +4 spaxels along the major axis indicated on the bottom panel on the right hand side of \autoref{fig:adamet_fiducial} (4 spaxels mean $0.20^{\prime \prime}$ which is approximately the size of the region we can extract kinematics data from). For each Voronoi bin `V' (located at coordinates $x_{V,k}$ and $y_{V,k}$ with $0\leq k \leq 16$) we compute the angular distances to each point `P' on galaxy's major axis ($x_{p,i}$ and $y_{p,i} =0$ for $0\leq i \leq 4$; we chose to sample the data-points at distances of $0.1 \arcsec$ from one another, giving a total of five points on the galaxy's major axis): $d_{i,k}= \sqrt{y_{V,k}^{2} + \left ( x_{V,k} - x_{p,i} \right )^{2}}$. Then for each constructed point on the galaxy's major axis we define a kinematic rotational velocity $V_{\rm kin,i}$ and velocity dispersion $\sigma_{kin,i}$ such that:

{\large
\begin{equation}
V_{\rm kin,i} = \left(\sum \limits_{k=0}^{16} V_{k} / d_{i,k}^{2}\right)\bigg /\left(\sum \limits_{k=0}^{16} 1/d_{i,k}^{2}\right),
\label{eq:V_calculation_v_sigma}
\end{equation}
\begin{equation}
\sigma_{\rm kin,i} = \left(\sum \limits_{k=0}^{16} \sigma_{k} / d_{i,k}^{2}\right) \bigg/ \left(\sum \limits_{k=0}^{16} 1/d_{i,k}^{2}\right).
\label{eq:V_over_sigma_calculations}
\end{equation}
}
\noindent $V_{k}$ and $\sigma_{k}$ are the rotational velocity and the velocity dispersion (for the Voronoi bin `k') computed by our \textit{Fiducial} dynamical model for the Voronoi bin `k'. In this section, $V_{k}$ and $\sigma_{k}$ can be regarded as the PSF-deconvolved kinematics map values obtained within the \textit{Fiducial} dynamical model (\autoref{fig:profiles_fiducial-1}). 
These are retrieved by setting the keywords in \textit{JAM}: \textsc{analytic\_los} = \textit{False} and \textsc{interp} = \textit{False}.

Using the equations \ref{eq:V_calculation_v_sigma} and \ref{eq:V_over_sigma_calculations}, we obtain the radial profile data-points (the pink cross symbols on \autoref{fig:profiles_fiducial-1}) at various radial distances from the galaxy's centre along the galaxy's major axis (at the positions $R_{i}$ with respect to galaxy's observed centre). 
We then compute the values for $V_{\ast}$ and $\sigma_{\ast}$ at the galaxy's outskirts (so in our case at R = 0.2$\arcsec$ along the major axis). 
This procedure gives us the projected value of $\left(V/\sigma\right)_{\ast}$. In order to calculate the de-projected value we use equation (A.2) from \citet{califa2019and_deproj} and obtain $\left(V/\sigma\right)_{\ast}=0.76 \pm 0.16$. 

\subsubsection{Specific angular momentum parameter $\lambda_{\rm 2R_{\rm eff}}$}
\label{sec:spin_parameter}

Our resolution does not allow us to compute the spin parameter within $1 \ R_{\rm eff}$ directly.
We have only one Voronoi bin within $1 \ R_{\rm eff}$ (out of the 17 Voronoi bins) but 6 Voronoi bins within $2 \ R_{\rm eff}$ of the galaxy's centre. The radial distances of the Voronoi bin ``k'' are $R_{ k} = \sqrt{x_{\rm rot,k}^{2} + y_{\rm rot,k}^{2}}$ where $x_{\rm rot}$ and $y_{\rm rot}$ are the Voronoi bins coordinates in the rotated system. We do not correct for inclination at this step yet as we will do the de-projection calculations at a later step. The spin parameter within $2 \ R_{\rm eff}$ can be computed by summing over the Voronoi bins within the said aperture. For the resolved stellar kinematics maps of the galaxy, $V_{k}$ and $\sigma_{k}$ we use 
the output stellar kinematics maps from the \textit{JAM} fiducial dynamical model. Following the definition from \citet{emsellem2007_lambda_def}:

{\large
\begin{equation}
\lambda_{2  R_{\rm eff}} = \frac{\sum \limits_{k}^{R_{k} < 2R_{\rm eff}} F_{k} R_{k} \left | V_{k} \right |}{\sum \limits_{k}^{R_{k} < 2R_{\rm eff}} F_{k} R_{k} \sqrt{V_{k}^{2}+\sigma_{k}^{2}}}.
\label{eq:lambda_2reff_equation}
\end{equation}
}

\noindent In this equation $F_{k}$ represents the light flux at the position of the Voronoi bin. This light flux distribution map can be obtained as a direct output of our \textit{JAM} dynamical model, but an alternative method is to simply add up all the signal from the pixels contained within each Voronoi bin. We chose to use the former method because it ensures better consistency with the input light maps that were used for \textit{JAM}. However, in order to estimate the uncertainties on the Voronoi bin fluxes, $F_{k}$, we proceeded by first creating a pseudo-NIRCam F200W band image of our target in order to mimic the photometric data from which the light maps inputs for \textit{JAM} were constructed. Each Voronoi bin spectrum was convolved with the transmission curve $\mathcal{T}\left(\lambda\right)$ of the F200W band\footnote{Information about the NIRCam filters can be found \href{https://jwst-docs.stsci.edu/jwst-near-infrared-camera/nircam-instrumentation/nircam-filters\#NIRCamFilters-tablenotes/}{here}.} such that for any wavelength slice $\lambda_{j}$:

\begin{equation}
F_{\rm bin \, k,pseudo}= \sum \limits_{j}^{} F_{\rm bin \, k} \left(\lambda_{j}\right) \times w \left(\lambda_{j}\right),
\label{eq:flux_pseudo}
\end{equation}

\noindent with the wavelength slices weights defined as $w \left(\lambda_{j}\right)= \mathcal{T}\left(\lambda_{j}\right)/ \sum \limits_{i}^{} \mathcal{T} \left(\lambda_{i}\right)$. The noise values associated with the pseudo-F200W images are calculated using:

\begin{equation}
N_{\rm bin \ k,pseudo}= \sqrt{\sum \limits_{j}^{} N_{\rm bin \, k}^{2} \left(\lambda_{j}\right)\times w^{2} \left(\lambda_{j}\right)}.
\label{eq:noise_pseudo}
\end{equation}

\noindent As a result, we will approximate the uncertainties on the \textit{JAM} Voronoi bin fluxes as $N_{\rm bin \ k,real} = F_{k}\times N_{\rm bin \ k,pseudo}/F_{\rm bin \ k,pseudo}$ which we will use to propagate the uncertainties on $\lambda_{2R_{\rm eff}}$ from equation \ref{eq:lambda_2reff_equation}. Doing the calculations and the error propagations, we obtain the raw (not de-projected and not PSF corrected) $\lambda_{2R_{\rm eff}}=0.26 \pm 0.03$. In order to account for PSF effects, we use the empirical correction given by equations 9 and 16 from \citet{lambda_PSF_corr} and plugging in our values for observed ellipticity, S\'{e}rsic index, effective radius from \autoref{tab:best_fit_parameters_pysersic_svi} and $\sigma_{\rm PSF}=0.044 \arcsec$, we obtain $\lambda_{2R_{\rm eff}}  \left( \rm  PSF \ corrected\right)=0.82 \pm 0.10$. This still needs to be corrected for de-projection effects, for which we use equation (A.3) from \citet[][originally from \citealt{atlas3_emsellem2011}]{califa2019and_deproj} and we obtain $\lambda_{2R_{\rm eff}}  \left(\rm  PSF \ corrected, \ deprojected\right)=0.85 \pm 0.10$.


Next, we can convert from $\lambda_{2R_{\rm eff}}$ to $\lambda_{R_{\rm eff}}$ using equation 2 from \citet{califa2019and_deproj}: $\lambda_{2R_{\rm eff}} = \left (1.19 \pm 0.14 \right ) \lambda_{R_{\rm eff}}$. With this extrapolation, we would obtain: $\lambda_{R_{\rm eff}}=0.72 \pm 0.09$. This $\lambda_{R_{\rm eff}}$ value enables us to make comparisons with observations from the literature (see \autoref{sec: Discussion}). The criterion to define a galaxy as a slow rotator was proposed as the equation 19 of \citet{michele_2016_IFS}: $\lambda_{R_{\rm eff}} < 0.08 + \epsilon/4$. The intrinsic ellipticity of our galaxy is $\epsilon \approx 0.38$ in the \textit{Fiducial} dynamical model. 
This certainly places our galaxy well within the fast rotator regime. We can instead consider the alternative criterion from \citet{atlas3_emsellem2011}: $\lambda_{R_{\rm eff}} > 0.31 \sqrt{\epsilon}$ for $z\sim 0$ fast rotators. In this case, we reach the same conclusion that GS-9209 is definitely a fast rotator. This is further illustrated by calculating the value for $\left (V / \sigma \right )_{R_{\rm eff}} = 0.93 \pm 0.27$ (obtained from $\lambda_{R_{\rm eff}}$ by using equation 18 from \citealt{michele_2016_IFS}) is in agreement, within the uncertainties, with the value computed in Section \ref{sec:V_sigma}. 

\subsubsection{$\boldsymbol{\kappa_{\rm rot, \ast}}$}
\label{sec: kappa_rot}

We want to compute how much of the total kinetic energy budget of our galaxy goes into stellar ordered rotation. This parameter is defined using the equation 1 of \citet{sales2012}. In order to determine its value in our case, we use the \textsc{jam.axi.intr} routine which computes the intrinsic first and second velocity moments for an arbitrary axisymmetric galaxy model.

The first step is to compute the light and the dark matter components \textit{MGE} parametrisation. For a Gaussian with a peak surface brightness $I_{k}$, dispersion $\sigma_{k}$, and observed axial ratio $q_{k}^{\prime}$ and intrinsic axial ratio $q_{k}$ (see Section \ref{sec:mgefit} for details), the luminosity density is (equation 38 from \citealt{michele2020spherical}):

\begin{equation}
\nu_{k}=\frac{I_{k}q_{k}^{\prime}}{q_{k}\sigma_{k} \sqrt{2\pi}}.
\label{eq:density_MGE}
\end{equation}

\noindent In our case, for all Gaussians of the light component in our parametrisation $q_{k}^{\prime}=0.749$ and $q_{k} =q_{\rm intr}=0.62$. For the dark matter component, we fit the classic NFW density profile (equation \ref{eq:nfw_profile}) using \textsc{mge\_fit\_1d} in a similar way to how we fit the S\'{e}rsic profile in Section \ref{sec:mgefit} for the luminuous component. The resulting best-fit peak surface densities of the DM Gaussians are converted to volume densities using again the equation \ref{eq:density_MGE} and assuming a spherical dark matter halo so for all of these Gaussians $q_{\rm DM}^{\prime} = q_{\rm DM} =1$. We next use the \textsc{jam.mge.radial\_mass} function in order to calculate the amount of luminous mass within the effective radius in order to scale the dark matter component in such a way that $f_{\rm DM} \left ( <2R_{\rm eff}\right)$ matches the result from our \textit{Fiducial} dynamical model (14.5\%). Combining the \textit{MGE} Gaussians for these two components, we obtain the gravitational potential parametrisation for our dynamical model that will be passed as input to \textsc{jam.axi.intr} (together with the luminous component parametrisation alone). We assume the central BH mass to be $10^{8.8} \ M_{\odot}$ \citep{CarnallQuenching}, the inclination angle is $i= 57.8^{\circ}$ and $\beta=1-\sigma_{z}^{2}/\sigma_{R}^{2} = 0.286$ (both values are computed based on the results of the \textit{Fiducial} dynamical model). 

For our intrinsic dynamical model we use a grid in a cylindrical system of coordinates with R = 0 at the centre of the galaxy and z = 0 in the central plane of this galaxy. The points in the grid at which we evaluate the model span from $0.5 \ R_{\rm eff} < R < 20 \ R_{\rm eff}$ and from $-R_{\rm eff} \sqrt{q_{\rm intr}} < z < + R_{\rm eff} \sqrt{q_{\rm intr}}$ with 1000 equally spaced points in both R and z ranges. At each of these points, this model evaluates four quantities:

\begin{itemize}
\item $\sigma_{R,j}^{2}, \sigma_{z,j}^{2}, \sigma_{\phi,j}^{2}$, the squared velocity dispersions along the radial, vertical and azimuthal directions.
\item $\left \langle v_{\phi}^{2} \right \rangle_{j}$, the second velocity moment along the tangential direction which in turn gives us the mean velocity of rotation along the tangential direction $\left \langle v_{\phi} \right \rangle_{j}=\sqrt{\left \langle v_{\phi}^{2} \right \rangle_{j} - \sigma_{\phi,j}^{2} }$.

\end{itemize}

The equation 1 from \citet{sales2012} can thus be re-written as:

\begin{equation}
\kappa_{\rm rot, \ast} =\frac{\sum \limits_{j} \left ( m_{j} \left \langle v_{\phi}\right \rangle_{j}^{2} \right)}{\sum \limits_{j} \left [ m_{j} \left(\sigma_{\rm R,j}^{2}+\sigma_{\rm z,j}^{2} + \left \langle  v_{\phi}^{2} \right \rangle_{j} \right) \right]} .
\end{equation}

\noindent The mass weights at each point of the (R, z) grid have been computed using equation 13 from \citet{cappellari2008jam} for the luminous component \textit{MGE} parametrisation and its analogous (equation 15 from \citealt{cappellari2008jam}) for the dark matter component Gaussians and for each point on the grid $m_{j} = \nu_{\rm lum} \left (R_{j}, z_{j}  \right) + \rho_{\rm DM} \left ( R_{j}, z_{j}\right)$. The total luminosity of a Gaussian with peak surface brightness $I_{k}$, dispersion $\sigma_{k}$ and observed axial ratio $q_{k}^{\prime}$ is $L_{k} =2 \pi q_{k}^{\prime} \sigma_{k}^{2} I_{k}$ and the equation is completely analogous for the total mass of a DM Gaussian. Using our results for these quantities, we obtain $\kappa_{\rm rot, \ast} = 0.39 \pm 0.08$. 

\section{Discussion}
\label{sec: Discussion}

\subsection{Dynamical Structure of GS-9209}
\label{sec:structure_section_overall}

In this subsection, we discuss the properties of GS-9209 resulting from our dynamical model. We combine this information with the findings of our kinematics analysis in order to infer the most likely scenarios for the mass assembly history for this galaxy. In \citet{CarnallQuenching}, the authors justify, based on their model SFH, that $z\sim3-5$ MQGs have likely gone through a starburst/sub-millimetre galaxy (SMG) phase prior to the observation epoch. Therefore, we will now focus on exploring what the possible low redshift descendants of MQGs like GS-9209 might be.

\subsubsection{Dark Matter Content - Comparison with Local Observations}
\label{sec:DM_content}


In the corner plot of \autoref{fig:adamet_fiducial}, we estimate a dark matter mass fraction within $2 \, R_{\rm eff}$ of $14.5^{+6.0}_{-4.2} \%$. 
Because we assume a classical NFW dark matter density profile, this allows us to subsequently extrapolate the fraction of DM within one effective radius: $f_{\rm DM} \left(< R_{\rm eff}\right)=6.3^{+2.8}_{-1.7}\%$, although it is crucial to stress that this quantity is obtained indirectly, because the spatial resolution of our data do not allow us to directly measure this quantity or the precise shape of the inner DM profile. 

We compare this inferred low dark matter fraction that we found for GS-9209 (within $1 \ R_{\rm eff})$ with the values reported by a number of observational studies. \citet{atlas15} analyse a sample of local ETGs within the $\rm ATLAS^{\rm 3D}$ survey and conduct \textit{JAM} dynamical modelling of these galaxies, finding a median $ f_{\rm DM} \left ( < R_{\rm eff}\right) \sim 13\%$ for their entire sample, but only about $\sim 9\%$ for their most robustly modelled galaxies. \citet{dynpop5} have a 20 times larger sample of low redshift ETGs and conduct a similar \textit{JAM} dynamical modelling analysis, obtaining that the average $f_{\rm} \left ( < R_{\rm eff}\right) \approx 10\%$ for $\sigma_{e} > 10^{2.1} \ \rm \kms$. \citet{cappellari_shetty2015} also find a median $f_{\rm DM} \left (< R_{\rm eff} \right) = 9\%$ for a sample of massive galaxies ($10^{10} M_{\odot} < M_{\ast} < 10^{12} M_{\odot}$) at $z \sim 0.8$. For the higher redshift sample of $1.4<z<2.0$ massive quiescent galaxies of \citet{mendel_dm}, the average DM fraction within the effective radius is lower ($6.6\%$ on average) compared with their likely local descendants. This is in line with the predictions of simulations such as \citet{hilz2012dm}, who find that $f_{\rm DM}\left(<R_{\rm eff}\right)$ increases as galaxies build up their stellar mass via mergers throughout their evolution. Hence, $f_{\rm DM}\left(<R_{\rm eff}\right)$ is also predicted to increase with cosmic time. In the case of post-Cosmic Noon major mergers, the increase is mild and mainly driven by mixing processes leading to an intrinsic change in the spatial distribution of light and dark matter components \citep{hilz+12}. However, minor mergers will lead to a more substantial increase in $\rm f_{\rm DM} \left (<R_{\rm eff}\right)$ because in this case, the increase in stellar effective size is significant \citep{nipoti+2009} meaning that now $R_{\rm eff}$ probes more extended regions of the dark matter halo (hence with more dark matter mass). Possible colour gradients could also have an impact, since the outskirts of MQGs are dominated by younger stellar populations (assuming an inside-out quenching scenario) hence with lower $M_{\ast}/L$ ratios. This was proved to be true for GS-9209 \citep{paper_with_dust} with the galaxy's outskirts at $r>0.3\arcsec$ from the centre being younger than the central region at $r<0.3 \arcsec$. Since our kinematics data only extend up to $0.2 \arcsec$ and our dark matter fractions are measured within apertures lower than $0.2 \arcsec$, this means that the gradient discovered by \citet{paper_with_dust} was probed on larger spatial scales and our assumption of spatially uniform $M_{\ast}/L$ is still valid. 

We provide a comparison between the dark matter fraction of GS-9209, $f_{\rm DM} \left(<2 \, R_{\rm eff} \right)$, and the predictions of numerical simulations (focusing on the \textit{TNG50-1} simulation box) in Appendix \ref{sec:appendix_dm}.

\subsubsection{Extremely Compact Nature and Possible Descendants of GS-9209}

Compact stellar sizes were reported for individual MQGs \citep{CarnallQuenching,rubiesEGS49,rubies73quenched2025} but also for sample studies \citep{ito2024_compact, ji2024_compact,wright2024_compact}. It means that prior to quenching, the cold gas fuelling the extreme star formation episodes in the history of such galaxies was either channelled to the galaxy's innermost central part (this is in agreement with the findings of previous studies based on cosmological simulations, e.g. \citealt{compaction1,compaction2,will2025_paper2}) or that a compaction episode happened post-quenching: when the gas in the original stellar disc was consumed and the turbulent motions disappeared, the stellar orbits got shrinked as a result of dynamical friction. 
The rotation pattern of the stellar component observed in GS-9209 tells us that the stellar disc is not destroyed by the quenching mechanism (at least not immediately). This seems to be in agreement with the inside-out quenching scenario \citep{tacchella2015_bulges} and it tells us that the formation of spheroid components takes place on a much longer timescale compared to quenching in the case of MQGs in the early Universe. 

The presence of a rotational pattern in the $V_{\ast}$ resolved map of this galaxy further indicates that this galaxy assembled most of its stellar mass, not by dry major mergers but more likely via secular accretion of cold gas possibly channelled along the cosmic web filaments (which ensures transportation of the gas deeper towards the central regions of massive halos). This picture is, however, challenged by the findings of \citet{blowing_out} who argue in the favour of the hypothesis that high-redshift MQGs may have initially been local nodes of the cosmic web residing in underdense environments. This means that the local filamentary structure is unable to efficiently accumulate mass at the node, leading to an extreme episode of isotropic, spherical mass accretion onto the node which is followed by a strong starburst phase. In any case, the subsequent quenching is likely the result of the action of AGN ejective feedback driving large outflows of cold gas, which is rapidly expelled out of the galaxy's circumgalactic medium. Preventative feedback causes the heating of the gas in the halo, which ensures the long-term quiescence of galaxies such as GS-9209 or ZF-UDS-7329 \citep{ZFUDS7329}. 

Gas-rich mergers (either major or minor) must also take place (likely before the quenching event) in the case of GS-9209, since this galaxy is somewhat rounder compared to both the average local and high-z pure discs. \citet{lagos+18_another} show that gas-rich mergers typically produce small changes in a galaxy's stellar specific angular momentum and, in some cases, it might even cause a galaxy to spin-up. Dry major mergers, however, are less plausible since they would distort or even destroy the stellar rotation pattern that we observe. This seems to be in agreement with the findings of \citet{david2025mergers} who determine that major mergers contribute less than 15\% to galaxies' stellar mass growth (in a redshift range $3<z<9$ and for a sample with stellar masses $8<\log \left (M_{\ast}/M_{\odot} \right)<10$). Such galaxies could be the progenitors of MQGs like GS-9209. Numerical simulations (e.g. \citealt{sandro_paper2019}) reveal that ex-situ processes become the dominant source of mass growth in the case of massive $M_{\ast}>10^{11} \ M_{\odot}$ galaxies only at lower redshifts $z<2$. \citet{qiao45to115} find that at the redshift of GS-9209, the merger rate per galaxy is about $2 \ \rm Gyr^{-1}$ on average. \citet{CarnallQuenching} calculate that the age of this galaxy at the observation time (i.e. $t_{\rm obs} - t_{\rm form}$) is 400-500 Myr. Thus, all these results support the possibility that GS-9209 has not undergone a significant major merger throughout its history prior to z = 4.66. However, any major mergers in the subsequent evolution of MQGs like GS-9209 will ultimately determine the possible descendants of early formed, early quenched massive galaxies.

Due to the high-z MQGs' compactness, many authors claimed them to be the progenitors of the compact, dense cores of Cosmic Noon and later ETGs \citep{CarnallQuenching,ji_core_progenitors,rubies73quenched2025}. Indeed, correlations between quenching and $M_{\rm bulge}$ exist \citep{atlas20,asa_bluck2014} and, furthermore, by $z \approx 2.5$ there is clear evidence for massive bulges and galaxies with suppressed star formation, at least in their core regions \citep{tacchella2015_bulges}. However, it is likely that a fraction of high redshift compact MQGs  similar to GS-9209 may not be involved in low redshift major mergers. They might be the progenitors of a rare class of compact, massive early-type galaxies known as `relics'. An example of such a galaxy is NGC 1277 \citep{ngc1277first,ngc1277}, which has seen little change in terms of its stellar populations or structural properties within the last 12 Gyr. Further evidence that internal processes have the largest contribution to the mass assembly of NGC 1277 is represented by the discovery that this galaxy predominantly hosts metal-rich (red) globular clusters \citep{red_clusters_ngc1277}. The stellar populations of relic galaxies are believed to be formed during a single starburst episode in the history of this galaxy. \citet{red_clusters_ngc1277} find that the fraction of blue, metal-poor globular clusters in NGC 1277 (whose likely origin is accretion from less massive companion galaxies) is significantly smaller compared to galaxies of similar $M_{\ast}$ to NGC 1277. 


Regarding the dark matter content, NGC 1277 is even more baryon dominated ($f_{\rm DM} \left( < 5 \ R_{\rm eff}\right) < 5\%$) than GS-9209. Both NGC 1277 and GS-9209 are fast rotators, highly compact and strongly baryon dominated, indicating that the latter galaxy is the currently highest-redshift analogue of the local relics galaxy population. A possible link between high redshift MQGs and low redshift relics has been suggested by \citet{haartman_highz_relic}. \citet{ngc1277} suggest that the low dark-matter content of relics in the local Universe might be caused by processes taking place in early stages of galaxy formation. One possible explanation could be that, during the early stages of massive galaxy formation, baryons cool and accrete onto the central galaxy on a shorter timescale than dark matter. An alternative possibility is given by an extensive dark matter component stripping during the galaxy's infall into a rich galaxy cluster (NGC~1277 is part of the Perseus Cluster, but this does not apply to GS-9209 which is an isolated object). Finally, the low dark matter fraction could also be explained by an early proto-galactic collision that essentially left the galaxy heavily baryon dominated for its entire subsequent evolution. 

Interestingly, GS-9209 and NGC 1277 possess a key difference: the initial mass functions (IMFs) of their stellar populations. \citet{ngc1277} found an extremely bottom-heavy IMF with $\alpha = \left(M_{\ast}/L\right)/\left(M_{\ast}/L\right)_{\rm Salpeter} = 1.3-1.5$ for NGC 1277. \citet{other2relics} also found that two additional relic galaxies (Mrk 1216 and PGC-032873) have bottom-heavy IMFs. Indeed, \citet{maciata_relics} reported (figure 5 of their paper) that extreme relic galaxies are generally characterised by a bottom-heavy IMF. However, we find that GS-9209 is instead characterised by a standard IMF. This is because our fiducial dynamical model returned $\log \left( M_{\ast}/M_{\odot}\right) = 10.52 \pm 0.06 $, which is in excellent agreement with the value found by \citet{CarnallQuenching} ($10.58 \pm 0.02$) who performed a SED fitting of the galaxy's spectrum and assumed a Kroupa IMF \citep{kroupaIMF}. This may point to some differences in the formation and mass assembly processes of early massive quenched galaxies, compared to low-redshift relic galaxies. Hence, it is unclear whether the former category truly represents the progenitors of the latter.

Regardless, our findings represent the first dynamical measurement of the IMF at $z>3$, highlighting the potential of future spatially resolved surveys of high redshift MQGs.

\subsection{\texorpdfstring{$\boldsymbol{\lambda_{R_{\rm eff}}}$}{lambda(Re)} Comparison with Literature}
\label{sec:lambda_discussion}
\begin{figure*}
    \centering

    \includegraphics[width=0.935\linewidth]{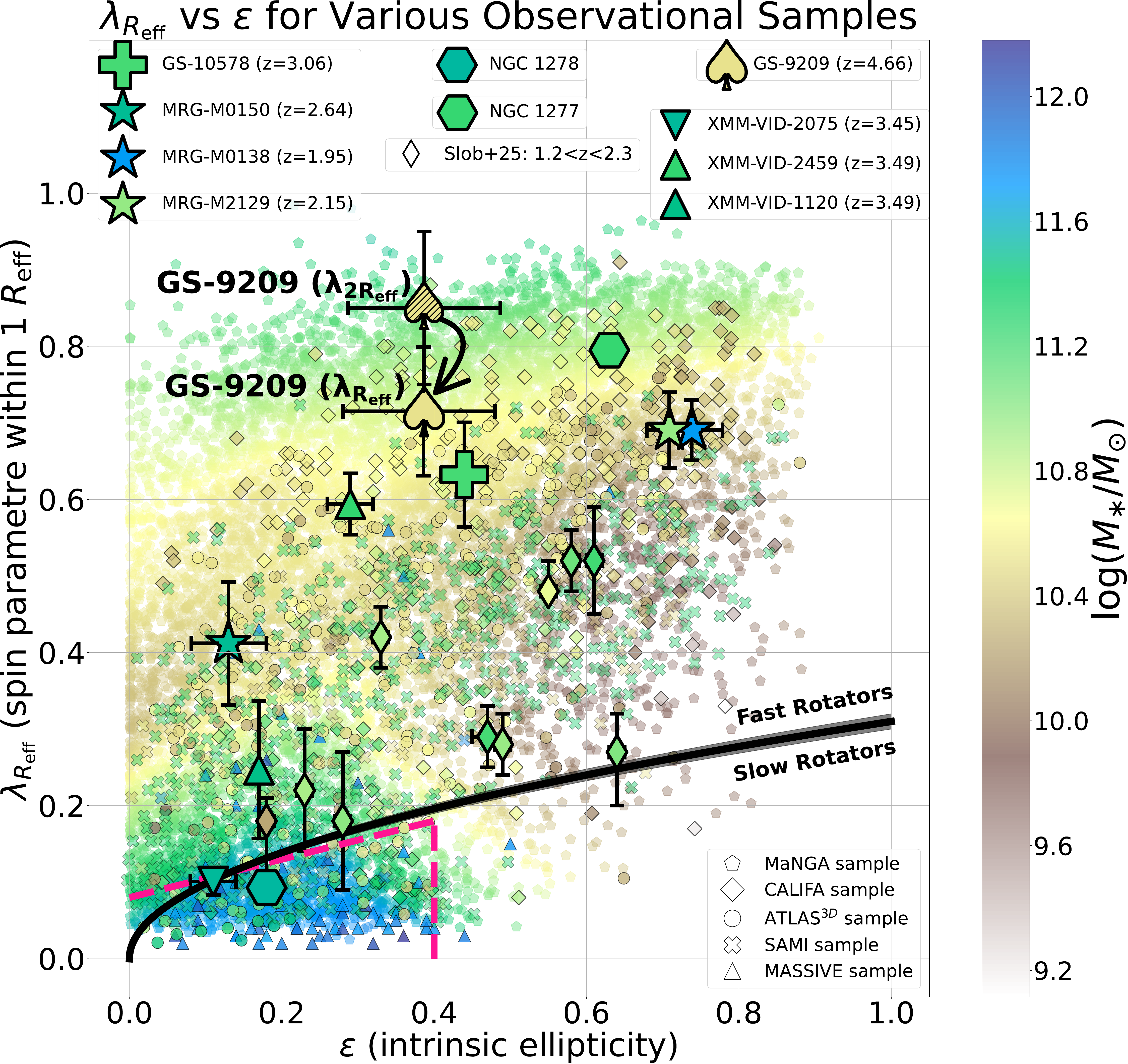}
             
    \caption{$\lambda_{R_{\rm eff}}$ against ellipticity, colour coded by $\log \left(\rm M_{\ast}/M_{\odot}\right)$. \textbf{The small symbols} are galaxies from various observational surveys that offer a comprehensive picture of the kinematic properties of local galaxies with $M_{\ast}>10^{9} \, M_{\odot}$. \textbf{Small pentagons} are $z<0.17$ galaxies from the MaNGA DynPop catalogue \citep{dynpop-1}. \textbf{Small diamonds} are local galaxies (0.005 $<$ z $<$ 0.03) from the \textit{CALIFA} survey \citep{califa2019and_deproj}. \textbf{Small circles} represent E and S0 type galaxies in the nearby Universe from $\mathrm{ATLAS^{3D}}$ survey \citep{atlas1}. For the purpose of their colour coding by stellar mass, we multiplied the r-band luminosity data from \citet{atlas15} with the stellar M/L ratio in the same band (from \citealt{atlas20}). Because these three data sets are large, we \textit{LOESS} smoothed (see \citealt{loess2d}) each of them. \textbf{X crosses} are z $<$ 0.095 galaxies from \textit{SAMI} survey \citep{vandesande_SAMI2017}. \textbf{Small triangles} represent extremely massive ($M_{\ast} > 10^{11.5} M_{\odot}$) local ETGs from the \textit{MASSIVE} sample \citep{massive5,massive7}. We overplot the positions on this diagram of 10 MQGs near Cosmic Noon ($1.2<z<2.3$ from \citealt{slob+25}; diamonds), three additional lensed MQGs at $z=1.93-2.64$ (MRG sample; \citealt{Newman2018_MRG}), two massive, dark matter-defficient galaxies, \textbf{NGC 1277 and NGC 1278} ($f_{\rm DM} \left(<5R_{\rm eff}\right)<0.05$ and $f_{\rm DM} \left(<1.5  R_{\rm eff}\right) = 0.14 \pm 0.04$ respectively) reported by \citet{ngc1277} and, finally, three galaxies at $z\sim3.5$ from the VISTA Deep Extragalactic Observations survey (VIDEO; \citealt{jarvis_video}), including the earliest slow rotator known, XMM-VID-2075 (\citealt{slow_rotator}; downwards pointing triangle). GS-10578 \citep{deugenio+2024} is the most similar fast rotator MQG to our target. The two fast/slow rotator boundaries follow the equations: $\lambda_{R_{\rm eff}} = 0.08 + \epsilon / 4$ \citep[for $\epsilon<0.4$; ][pink dashed line]{michele_2016_IFS} and $\lambda_{R_{\rm eff}} = 0.31 \cdot \sqrt{\epsilon}$ \citep[][black solid line]{atlas3_emsellem2011}. In this plot, we also illustrate the empirical relation $\lambda_{2R_{\rm eff}}=\left(1.19\pm0.14\right)\times\lambda_{R_{\rm eff}}$ \citep{califa2019and_deproj} that we used to estimate the observationally inaccessible $\lambda_{R_{\rm eff}}$ (due to spatial resolution limitations) based on our measurement of $\lambda_{2R_{\rm eff}}$. Admittedly, this empirical correction is not perfect, but it is accurate enough to preserve our conclusion that GS-9209 is unambiguously a fast rotator.}
    \label{fig:lambda_big_plot}

\end{figure*}

Our value of $\lambda_{2R_{\rm eff}} = 0.85 \pm 0.10$ gives $\lambda_{R_{\rm eff}}=0.72 \pm 0.09$ (as shown in Section \ref{sec:V_sigma}). 
We can study the properties of GS-9209 by comparing it with GS-10578 (another Massive Quiescent Galaxy, but at lower redshift, $z \sim 3.064$). As we can observe in Fig. \ref{fig:lambda_big_plot} these galaxies occupy a similar place in the $\lambda_{R_{\rm eff}}-\epsilon$ plane but they have quite different stellar masses. It is important to look in more detail at some potential key differences between the way these two galaxies assembled their mass and quenched star formation.

\citet{deugenio+2024} mention that GS-10578 had its main star formation episodes at epochs $z = 3.7 - 4.6$. The stellar mass of GS-10578 is approximately 5 times larger than that of GS-9209. Both galaxies retain high values of specific angular momentum which means that these galaxies formed most of their stellar mass \textit{in-situ} and that they likely did not experience significant gas-poor mergers (otherwise we would not observe a significant stellar rotation; e.g. \citealt{lagos+18_another}). If galaxies like GS-9209 were the direct progenitors of those like GS-10578, given the stellar mass differences between the two galaxies, and the difference in the ages of the Universe at these redshifts, this would imply an average stellar mass growth rate of $\dot{M}_{\ast} \geq \rm 200 \ M_{\odot} \ \rm yr^{-1}$. If this was fuelled solely by fresh cold gas accretion (major mergers would disturb the observed stellar disc motions), that new reservoir of cold gas would need to be converted into stars in order to match the observational constraints of low molecular hydrogen mass fraction in $z>3$ quiescent galaxies (e.g. \citealt{suzuki_lowgas,jan_net0}). Such high values of SFR are typically associated with high redshift sub-millimetre galaxies (SMGs) or high redshift dusty star forming galaxies \citep[DSFGs; ][]{carniani2013,gilli+14,tan+14_SMG,casey+19,sharda+19,riechers+20,rizzo+20,birkin+21,fraternali+21,rizzo+21,ikarashi+22,fengwu+23,parlanti+23,arribas+24_proto,fuentealba+24,yongda_zhu+24,sillassen+25}. Consequently, a plausible explanation is that $z>4$ SMGs and/or DSFGs could be the progenitors of MQGs like GS-9209 or GS-10578. Some galaxy formation and evolution simulations predict an evolutionary link between SMGs and MQGs \citep{lagos15_original}. 

\citet{minjungpark2024_outflows} analysed SFHs in Cosmic Noon MQGs and they report having found their sample to be split into three different sub-categories (figure 7 of that paper). In our case, the findings of \citet{minjungpark2024_outflows} resonate with the different star formation histories of GS-10578 (which is a recently formed, recently quenched galaxy) and GS-9209 (an early formed, early quenched galaxy). As a result, despite showing some similarities, galaxies such as GS-9209 are unlikely to be the direct progenitors of those like GS-10578. This underlines the idea that the likely mechanisms of mass assembly and quenching of massive quiescent galaxies were already in place at $z>5$ and, moreover, the same processes seem to occur in a similar way for widely different stellar masses.

Close similarity in terms of the position in the $\lambda_{R_{\rm eff}} - \epsilon$ plane can be observed in the case of the $z \sim 2$ galaxies MRG-M0138 and MRG-M2129. However, just as in the case of GS-10578 vs GS-9209, the stellar masses of these Cosmic Noon quiescent galaxies are also widely different: $M_{\ast} = 10^{10.96} M_{\odot}$ for MRG-M2129 and $M_{\ast} = 10^{11.69} M_{\odot}$ for MRG-M0138. In the case of these galaxies, the stellar rotation component (highlighted by the high ellipticity and high $\lambda_{R_{\rm eff}}$) has been preserved after quenching. In addition, the galaxies are observed (via lensing) at similar redshifts ($z \sim 2.15$ for MRG-M2129, \citealt{toft2017_MRG} and $z \sim 1.95$ for MRG-M0138, \citealt{Newman2018_MRG}) which, given the highly different stellar masses, argues for different star formation histories in these galaxies. Again, given the high $\lambda_{R_{\rm eff}}$, it is most likely that these galaxies built most of their stellar masses via secular cosmological gas accretion and \textit{in-situ} gas conversion into stars. Nevertheless, the quenching mechanism leading to the halting of star formation in such massive galaxies seem to be taking place in a similar way independent of $M_{\ast}$. It can be remarked that the MQGs at $z>1.5$ that we plotted in \autoref{fig:lambda_big_plot} and the relic galaxy NGC 1277 at $z=0$ are all fast-rotators. This is in very good agreement with the findings from the \textit{Magneticum Pathfinder} simulations. Figure 10 from \citet{blowing_out} highlights that at all redshifts $2.7<z<5.4$ tested by the authors, slow rotators quiescent galaxies represent a minority of the overall galaxy population.

\subsection{\texorpdfstring{$\boldsymbol{V/\sigma}$}{v/s} Comparison with Observations}
\label{sec: Vsigma_observations}

\begin{figure*}
    \centering
    \includegraphics[width=\textwidth]{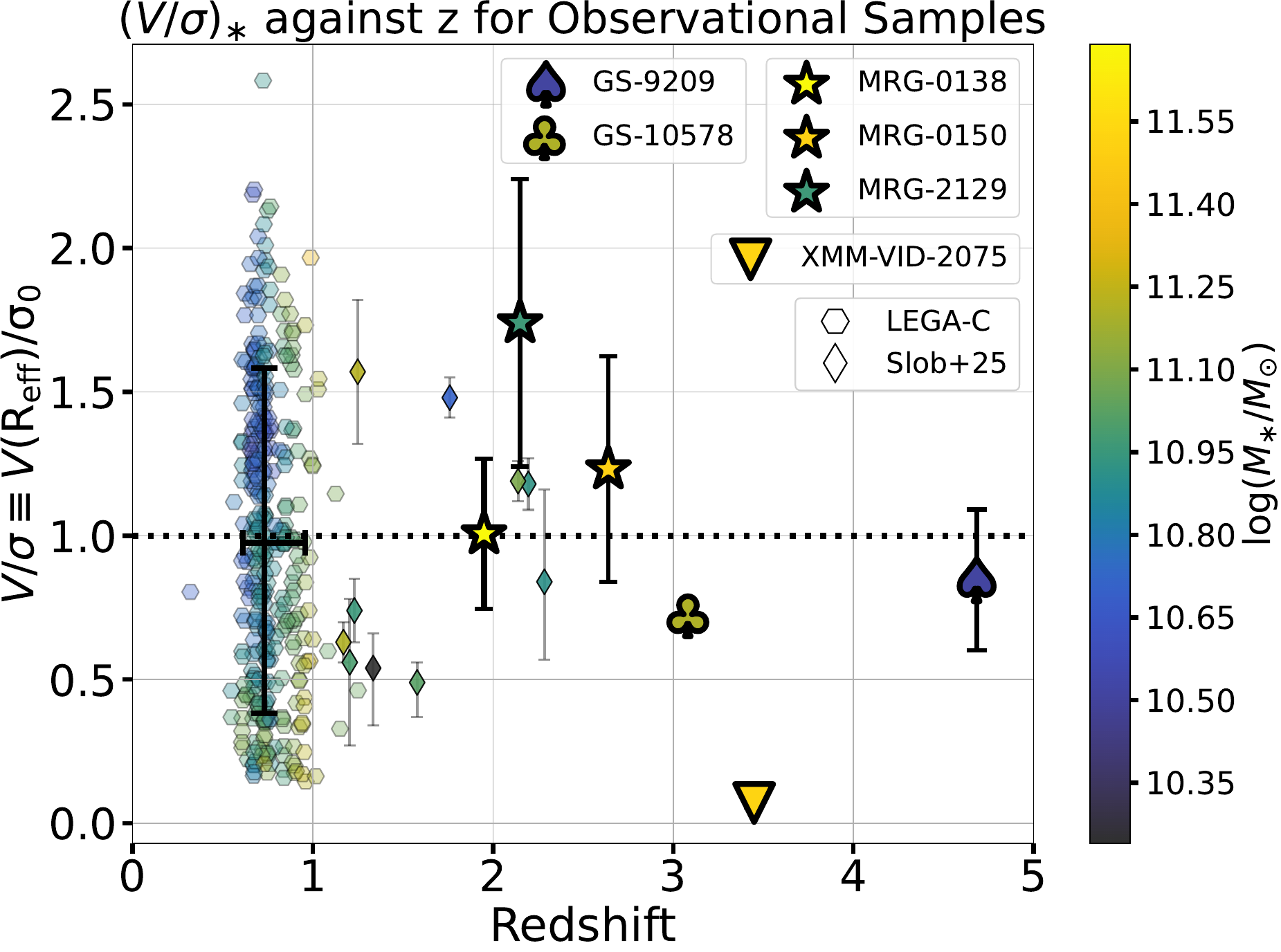}
    \caption{This figure illustrates the ratios of rotational velocity to velocity dispersion, $\left(V/\sigma\right)_{\ast}$ as traced by stellar kinematics of various targets at $0.5<z<5$. All targets are colour coded using the logarithm of the stellar mass. GS-9209 (presented in this work) is the highest-redshift object for which this measure of rotational support was determined. \textbf{The background hexagons} are $0.6<z<1$ massive quiescent galaxies whose stellar kinematics were determined with Jeans anisotropic models \citep{vanhoudt+2021}. The two thick error-bars indicate the median redshift and $V_{R_{\rm eff}}/\sigma_{0}$, together with their $\left[5;\ 95\right]\%$ and $\left[16; \ 84\right]\%$ ranges, respectively. \textbf{The ten small diamonds} are a sample of massive quiescent galaxies near Cosmic Noon ($1.2<z<2.3$) from the \textit{SUSPENSE} survey \citep{slob+25}. We also overplot the three lensed MQGs at $1.95<z<2.64$ from \citet{Newman2018_MRG}, the recently discovered massive slow rotator at $z=3.45$ (XMM-VID-2075; \citealt{slow_rotator}) and the GS-10578 \citep{deugenio+2024} fast-rotator, with the most similar kinematic properties to GS-9209. The existence of XMM-VID-2075 slow-rotator is surprising, given that most of the other $z>1$ MQGs are fast rotators. The existing observational data is too scarce to constrain an evolutionary trend with redshift, but excluding XMM-VID-2075, the other data-points shown here seem to indicate that the stellar $V/\sigma$ ratio does not evolve sigificantly across cosmic time.}
    \label{fig:vsigma_stellar}
\end{figure*}

In \autoref{fig:vsigma_stellar}, we plot the evolution of the ratio between stellar rotational velocity and stellar velocity dispersion ($V/\sigma \equiv \rm V\left(R_{\rm eff}\right)/\sigma_{0}$) based various observational samples. We present the quiescent galaxies in the \textit{LEGA-C} and \textit{SUSPENSE} samples, together with the six MQGs at $z>1.5$ which have available stellar kinematics data. 
In cases where $V \left(R_{\rm eff}\right)/\sigma_{0}$ is not directly reported, we compute this ratio as follows. For five of our individual MQGs highlighted in the plot (GS-9209, GS-10578 from \citealt{deugenio+2024}, MRG-0138, MRG-0150 and MRG-2129 from \citealt{Newman2018_MRG}), we use their radial profiles of the velocity dispersion $\sigma \left(\rm R\right)$ to determine the ratio $\sigma \left(R=0\right)/\sigma \left(R=R_{\rm eff}\right)$. In the case of the slow rotator XMM-VID-2075, \citet{slow_rotator} report $\sigma_{0}\sim 500 \ \rm km \ s^{-1}$ and $\sigma_{R_{\rm eff}}=387 \pm 22 \ \rm km \ s^{-1}$, from which we obtain a value of $1.29 \pm 0.07$ for this ratio. We then estimate the values of $\left(V/\sigma\right)_{R_{\rm eff}}$ based on the measured $\lambda_{R_{\rm eff}}$ parameters using equation 18 from \citealt{michele_2016_IFS}. We then multiplied $\left(V/\sigma \right)_{R_{\rm eff}}$ by $\sigma \left(\rm R_{\rm eff}\right)/\sigma_{0}$. The existence of a slow rotator MQG at $z\sim3.44$ could be potentially explained as the outcome of a major merger event which strongly disrupted a previously existing stellar disc. This would also be consistent with the relatively high observed axial ratio of that object ($q^{\prime}\sim0.89$; \citealt{slow_rotator}). However, an alternative hypothesis of an isotropic gas infall has been proposed by \citet{chandro-gomez+2025}. 


For the three lensed massive quiescent targets from \citet{Newman2018_MRG}, their effective radii are given in \citet{newman_paper1} in the case of MRG-2129 and MRG-0150 and in \citet{newman2025} for MRG-0138. We have assumed that for these three MQGs $\sigma_{0} \approx \sigma \left(R=0 \right)$. This approximation is true in the case of GS-9209 ($\sigma_{\ast}^{\prime} \approx 239 \pm 34$ km/s, see Section \ref{sec:pPXF_fitting} while $\sigma^{\prime} \left(\rm R=0 \right) \approx 230 -240$ km/s, see Fig. \ref{fig:profiles_fiducial-1}). The numerical uncertainty of the aperture-integrated velocity dispersion that we measure is larger than the difference between $\sigma_{\ast}^{\prime}$ and $\sigma^{\prime} \left(R=0 \right)$ by an order of magnitude.

In the case of quiescent galaxy samples with stellar kinematics, there is no significant trend between $\left(V/\sigma\right)_{\ast}$ and redshift (see \autoref{fig:vsigma_stellar}). 
However, this sample of $z>2$ MQGs, together with the quenched galaxies from \textit{LEGA-C} and \textit{SUSPENSE}, is not large enough to obtain a statistically significant conclusion. Within the \textit{LEGA-C} sample, lower-mass quiescent galaxies have more rotational support. This is in contrast with the findings of $z<3$ ionised gas kinematics surveys targetting star-forming galaxies \citep{wisnioski2015,simons_SIGMA+16,forster_schreiber2018_sins,wisnioski2019,genzel+20}; in these cases, the galaxies with the highest $V/\sigma$ ratios are also very massive \citep{wisnioski2019,lola2025}. This highlights the different behaviour of distinct kinematics tracers (e.g. star and warm ionised gas, in this case), which furthermore implies that $\rm \left(V/\sigma\right)_{\ast}$ and $\rm \left(V/\sigma\right)_{\ast}$ cannot be compared directly. This is because ionised gas typically traces a more dynamically active interstellar medium, while the stars in fast-rotators MQGs like GS-9209 or GS-10578 have reached a steady-state ordered rotation.  

\section{Conclusions}
\label{sec: Conclusions}

In this paper, we used ultra-deep (14.7 hours) high-resolution IFU \textit{JWST}/NIRSpec data
with the G235H grating to conduct a detailed analysis of the structural, kinematic and dynamic properties of the $z \sim 4.66$ massive quiescent galaxy GS-9209.

\begin{itemize}
\item From the integrated aperture spectrum, we measure a stellar velocity dispersion of $\sigma_{\ast}^{\prime} = 239 \pm 34$ \kms in agreement with the value found by \citet{CarnallQuenching} ($247 \pm 16$ \kms, obtained using medium-resolution spectroscopy).
\item We showed the resolved maps $V_{\ast}$ and $\sigma_{\ast}$, extending out to $\sim 6 \ \rm R_{\rm eff} \sim 1.3 \ \rm kpc$. The stellar velocity map shows clear evidence for ordered rotation,  which makes GS-9209 the highest-redshift galaxy for which such a feature is probed via integral field spectroscopy.
\item We find that GS-9209 is rotation supported, having  a high (PSF-corrected) $\lambda_{2R_{\rm eff}} = 0.85 \pm 0.10$, from which we can extrapolate that $\lambda_{R_{\rm eff}} =0.72 \pm 0.09$, comparable to other quiescent galaxies at $z=2\text{--}3$. 
This value of $\lambda_{R_\mathrm{\rm eff}}$ is equivalent to a $\left (V/\sigma \right)_{R_{\rm eff}} = 0.93 \pm 0.27$. This means that the mechanism responsible for quenching this galaxy still preserved the rotating stellar disc. GS-9209 is a fast rotator similar to other intermediate and high-z massive quiescent galaxies \citep{Newman2018_MRG,deugenio+2024}. Furthermore, the high degree of rotational support suggests that the kinematic transformation between fast and slow rotators observed at lower redshifts occurs only after quenching. This is in agreement with the results obtained from numerical simulations: quenching preceedes kinematic transformation in galaxies that become slow rotators at $z\sim0$ \citep{sandro_paper2019,lagos+22}.
\item We then fit our fiducial dynamical model (with a classical NFW DM profile and a BH whose mass takes a Gaussian prior motivated by virial scaling relations based on its broad-line H$\alpha$ emission properties). The model assumes a cylindrical alignment of the velocity ellipsoid. With our fiducial dynamical model, we obtain the currently highest redshift measurement of the dark matter content of a galaxy based on its stellar kinematics: $f_{\rm DM} \left ( <2R_{\rm eff} \right ) = 14.5^{+6.0}_{-4.2} \%$. Given the compact nature of the source (compared to e.g. its host dark matter halo) the low DM fraction is in line with the expectations. Other important parameters that come out from the dynamical modelling are the intrinsic axial ratio and the radial anisotropy. These are crucial in our de-projection calculations for the kinematics parameters.
\item Also from our fiducial dynamical model, we infer a $\left(M_{\ast}/L \right) = 0.10 \pm 0.02 \ M_{\odot}/L_{\odot}$. This represents the first such measurement at $z>2$ and it is consistent with a standard, Milky-Way like IMF from stellar population modelling. The agreement with the value inferred from SED modelling \citep{CarnallQuenching} provides independent confirmation
on the large stellar mass of early quiescent galaxies.
\item The low dark matter content of GS-9209 and its high spin parameter make it similar to the $z \sim 0$ population of relic galaxies (e.g. \citealt{ngc1277}). This remarkable category of galaxies has undergone little evolution within a time interval of $\sim 10$ Gyr or more. However, one possible key difference between high redshift MQGs and low redshift relics is the difference between their IMFs. Further dynamical studies on high z MQGs would reveal their IMFs, which could confirm or refute the possibility of them being the progenitors of low redshift relic galaxies.
\item We compared the value of $\lambda_{R_{\rm eff}}$ of GS-9209 (corrected for projection and PSF effects) to independent observational surveys in the $z < 1$ Universe but also to various $z>2$ particular MQGs for which this parameter was calculated. For virtually all intermediate and high-$z$ MQGs (except for one recently discovered high-redshift slow rotator) the positions on the $\lambda_{R_{\rm eff}} - \epsilon$ plane indicate that they are fast rotators. This is a strong evidence that the star formation quenching and the destruction of the stellar disc likely happen in this precise order. The causality relation between these two important milestones in a galaxy's evolution is yet to be fully understood. For the general galaxy population, as they build up their stellar mass, the amount of rotational support diminishes. This is in good agreement with the current galaxy evolution theories (e.g. \citealt{kinematics_magneticum} analyse the evolution of $\lambda_{R_{\rm eff}}-\epsilon$ plane at $z<2$ in the \textsc{Magneticum} simulations, reporting an increased fraction of slow rotators). This is correlated with the formation and growth of central bulges and dispersion dominated spheroidal sub-structures.
\end{itemize}

The difficulty encountered by multiple simulations to reproduce the observed number density of MQGs and their physical properties (the most problematic one being the compact size) points to key differences between the processes governing the evolution of galaxies within $t<1$ Gyr and the ones implemented in theoretical models. By exploring these differences, we can improve our understanding of the complex physical processes shaping the SFHs of high-redshift galaxies, their mass assembly histories and the spatial distribution of baryons. In this direction, our work highlights the crucial ability of stellar kinematics data to provide constraints for cosmological simulations. From an observational point of view, studying quiescent galaxies enables us to measure their dark matter content without the complication of a possible large gas fraction such as in star-forming galaxies. Constructing a sample of high redshift quenched galaxies with spatially resolved stellar kinematics would provide tighter constraints on current theoretical models. 

In the absence of galactic outflows, this galaxy could be able to rejuvenate in the future via merger induced starbursts, for instance. The analysis of spatially resolved outflow properties in high-$z$ MQGs could give more clear insights into their possible descendants. 
Another possible study would also involve FIR/sub-mm observations of this galaxy with ALMA, having the primary goal of investigating the amount of cold molecular gas and the state of that gas. This can be done with the $\left [\rm C \textsc{ii} \right ]\lambda 158 \mu m$ fine structure emission line or with CO emission lines, which are both tracers of molecular gas even for quiescent galaxies \citep{deugenio2023_cII_coldgas}. Such studies can be extended to samples of multiple MQGs at $z>2$, which would allow us to determine both the cold gas content and the gas-consumption histories of MQGs more accurately. This is because high-redshift MQGs are usually quenched close (less than 1 Gyr) to the observation time, so passive evolution effects are minimal, which puts more stringent constraints on the nature of the quenching mechanism. Such a sample could be compared with the results of numerical simulations (e.g \citealt{lorenzon+25}) and thus provide an extremely valuable step forward in constraining the physics implemented in the next generation of galaxy evolution simulations.

\section*{Acknowledgements}
We are grateful to the referee for insightful and constructive feedback that substantially improved the quality of this manuscript. RGP acknowledges funding support from a STFC PhD studentship. RGP, FDE, RM and JS acknowledge support by the Science and Technology Facilities Council (STFC), by the ERC through Advanced Grant 695671 ``QUENCH'', and by the UKRI Frontier Research grant RISEandFALL. RM also acknowledges funding from a research professorship from the Royal Society. ST acknowledges support from the Royal Society Research Grant G125142. SC acknowledges support by European Union's HE ERC Starting Grant No. 101040227 - WINGS. AJB acknowledges funding from the ``FirstGalaxies'' Advanced Grant from the European Research Council (ERC) under the European Union's Horizon 2020 research and innovation program (Grant Agreement No. 789056). SA and MP acknowledge grant PID2021-127718NB-I00 funded by the Spanish Ministry of Science and Innovation/State Agency of Research (MICIN/AEI/ 10.13039/501100011033); MP also acknowledges the grant RYC2023-044853-I, funded by  MICIU/AEI/10.13039/501100011033 and European Social Fund Plus (FSE+). We thank William Baker and Yuki Isobe for insightful discussions that led to improvements of the original manuscripts and paper drafts.

\section*{Data Availability}
 
This work is based on observations conducted with the NASA/ESA/CSA James Webb Space Telescope. The data were obtained as part of the JWST program ID 3659 (Cycle 2, PI: Francesco D'Eugenio). These data are available from \href{https://mast.stsci.edu/portal/Mashup/Clients/Mast/Portal.html}{Mikulski Archive for Space Telescopes} at the Space Telescope Science Institute, which is operated by the Association of Universities for Research in Astronomy , Inc., under NASA contract NAS 5-03127 for JWST. The fully reduced datacube is available upon reasonable request. The photometry data in the JWST/NIRCam F200W band are available from the public data release page of the \href{https://archive.stsci.edu/hlsp/jades}{JADES survey on MAST}.

\bibliographystyle{mnras}
\bibliography{refs} 


\vspace{-0.4cm}
\section*{}
\noindent
\textit{\small{
$^{1}$ Kavli Institute for Cosmology, University of Cambridge, Madingley Road, Cambridge CB3 0HA, UK \\
$^{2}$ Cavendish Laboratory – Astrophysics Group, University of Cambridge, 19 JJ Thomson Avenue, Cambridge CB3 0HE, UK \\
$^{3}$ Department of Physics and Astronomy, University College London, Gower Street, London WC1E 6BT, UK \\
$^{4}$ Sub-Department of Astrophysics, Department of Physics, University of Oxford, Denys Wilkinson Building, Keble Road, Oxford, OX1 3RH, UK \\
$^{5}$ International Centre for Radio Astronomy Research, The University of Western Australia, 35 Stirling Highway, Crawley,
Western Australia 6009, Australia \\
$^{6}$ ARC Centre for All-Sky Astrophysics in 3 Dimensions (ASTRO 3D) \\
$^{7}$ Max-Planck-Institut f{\"u}r extraterrestrische Physik (MPE), Gie\ss{}enbachstra\ss{}e 1, 85748 Garching, Germany \\ 
$^{8}$ INAF - Osservatorio Astrofisico di Arcetri, Largo E. Fermi 5, I-50125, Firenze, Italy \\
$^{9}$ Centro de Astrobiolog\'ia (CAB), CSIC-INTA, Cra. de Ajalvir Km. 4, 28850- Torrej\'on de Ardoz, Madrid, Spain \\
$^{10}$ Sterrenkundig Observatorium, Universiteit Gent, Krijgslaan 281 S9, 9000 Gent, Belgium \\  
$^{11}$ NSF National Optical-Infrared Astronomy Research Laboratory, 950 North Cherry Avenue, Tucson, AZ 85719, USA \\
$^{12}$ Steward Observatory, University of Arizona, 933 North Cherry Avenue, Tucson, AZ 85719, USA \\
$^{13}$ Max-Planck -Institut für Astronomie, K{\"o}nigstuhl 17, Heidelberg, Germany  \\
$^{14}$ Max Planck Institute for Astrophysics, Karl-Schwarzschild-Str. 1, 85741 Garching bei München, Germany \\
$^{15}$ Ludwig-Maximilians-Universität München, Geschwister-Scholl-Platz 1, 80539 München, Germany \\
$^{16}$ Department of Astronomy, University of Wisconsin-Madison, Madison, WI 53706, USA \\
$^{17}$ Institute for Astronomy, University of Edinburgh, Royal Observatory, Edinburgh, EH9 3HJ, UK \\
$^{18}$ Sorbonne Universit\'e, CNRS, UMR 7095, Institut d'Astrophysique de Paris, 98 bis bd Arago, 75014 Paris, France \\
$^{19}$ Scuola Normale Superiore, Piazza dei Cavalieri 7, I-56126 Pisa, Italy \\
}}
\FloatBarrier
\appendix

\section{Best-fit parameters from \textsc{PySersic} Photometry Fitting}
\label{sec: appendix 1}

These tables present the best-fit parameters that we obtain from \textsc{PySersic} photometry fitting of the galaxy image in NIRCam's F200W filter. The first table gives the parameters obtained using the Laplace sampling method of retrieving the best-fit parameters from the chains' samples. In general, the SVI flow method is more reliable and accurate so we chose to use these values in our subsequent analysis. For the data reported in \autoref{tab:best_fit_parameters_pysersic_laplace}, we use a prior assumption of a simple S\'ersic profile across all the galaxy. 

\begin{table}
\centering
\renewcommand{\arraystretch}{1.35} 
\setlength{\tabcolsep}{14pt}      
\captionsetup{justification=centering}
\caption{Best Fit Parameters for \textsc{PySersic} F200W Image - Laplace sampling method}

\begin{tabular}{ c  c }
\hline
\textbf{parameter} & \textbf{Value} \\ 
\hline
observed ellipticity & $0.252 \pm 0.015$ \\ 

S\'ersic index & $2.958 \pm 0.140$ \\ 

$R_{\rm eff} $ (arcsec) & $0.0345 \pm 0.0007$ \\ 

$\theta$ (deg) & $174.3 \pm 7.0$ \\ 

\hline
\end{tabular}

\label{tab:best_fit_parameters_pysersic_laplace}
\end{table}











\section{JAM Dynamical Model Test for the central Black Hole mass}
\label{sec:appendix_BHMFL}

\begin{figure*}
    \centering
    \includegraphics[width=\linewidth]{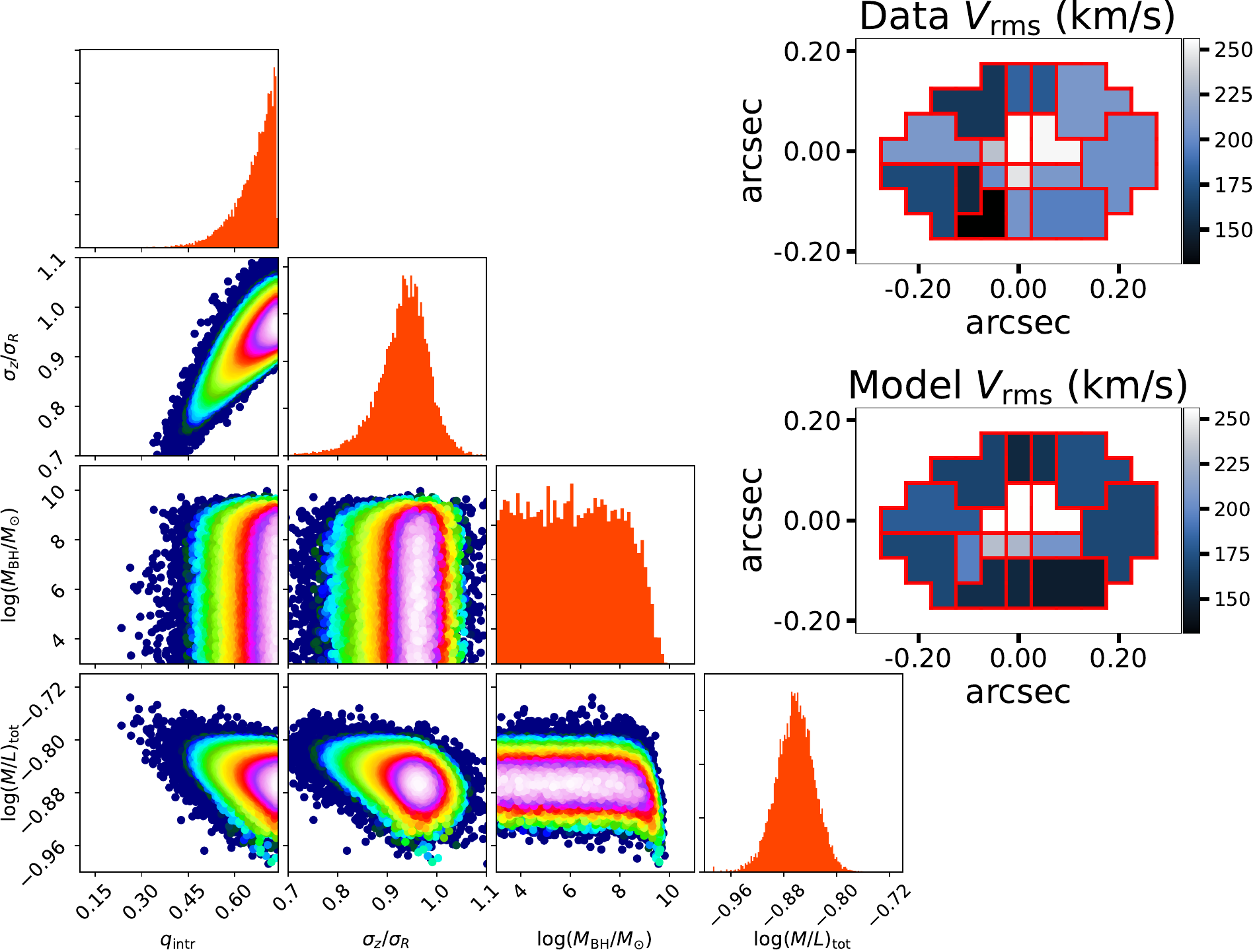}

    \caption{Outputs of the model \textit{Alternative-1}, assuming that we only have a central BH, which, together with the stellar mass itself, are responsible for producing the gravitational potential (hence the stellar kinematics). Because the galaxy is quiescent, we can safely neglect the mass of interstellar gas (and dust). This is a Black-Hole \& Mass-Follows-Light (BH+MFL) model. The best-fit parameters predicted by this model are given in Table \ref{tab:table_just_alternative1}. The two panels on the right hand side show the input vs the output map of $V_{\rm rms}$ predicted by this model. In this plot we denote $\log \left(M/L\right)_{\rm tot} \equiv \log \left[\left(M/L\right)_{\rm tot} /\left(M_{\odot}/L_{\odot}\right) \right]$. This model has reduced $\chi_{r}^2 = 2.7$.}

    \label{fig:adamet_BH_MFL}
\end{figure*}

This alternative dynamical model (\textit{Alternative-1}) is a simple mass-follows-light (MFL) model (i.e. no dark matter component) with an additional central BH mass ($M_{\rm BH}$). As we can remark from \autoref{fig:adamet_BH_MFL}, we cannot pose tight constraints on $M_{\rm BH}$ but we are able to reject central black holes with masses: $\rm M_{\rm BH} > 10^{9.27} \ \rm M_{\odot}$ at $2\sigma$ level (97.7\%) and $M_{\rm BH} > 10^{9.76} \ \rm M_{\odot}$ at $3\sigma$ level (99.9\%).

Using $\mathrm{M_{\rm BH} = 10^{8.8} \ M_{\odot}}$ and $\sigma_{\ast}^{\prime} = 239 \ \kms$, the radius of the sphere of influence of the black hole is about $R_{\rm i,BH} = \rm GM_{\rm BH}/\sigma_{\ast}^{\prime2} = 50 \ \rm pc \equiv$ 7.7 mas, 6.5 times below the pixel resolution of our IFU data ($0.05\arcsec$). This angular resolution limitation is the main reason for which we can only infer an upper black hole mass for GS-9209. The inferred stellar mass is $\mathrm{\log \left (M_{\ast} / M_{\odot} \right ) = 10.61 \pm 0.03}$, in remarkable agreement with the value obtained from the \textit{Fiducial} dynamical model (see Sec. \ref{sec:dynamical_modelling}).

\begin{table}
\centering
\renewcommand{\arraystretch}{1.35} 
\setlength{\tabcolsep}{11pt}    
\caption{Marginalized posterior probabilities on the free parameters of the \textit{Alternative-1} model (BH allowed to be fit freely and Mass Follows Light). The values reported are the median and 16\textsuperscript{th}--84\textsuperscript{th} percentile probability range.}

\begin{tabular}{ c c c}
\hline
Parameter & Value & Model \\ 
\hline
$q_{\rm intr} $ & $0.67 \pm 0.06$ & \textit{Alternative-1}  \\ 
$\sigma_{z} / \sigma_{R}$ & $0.94 \pm 0.05$  & \textit{Alternative-1} \\ 
$\log M_{\rm BH}/M_{\odot}$ & < 9.27 ($2\sigma$) & \textit{Alternative-1} \\ 
$\log \left[ \left ( M/L \right )_{\rm tot}/\left(M_{\odot}/L_{\odot}\right) \right] $ & $-0.91 \pm 0.03$ & \textit{Alternative-1}  \\ 

\hline

\end{tabular}
\label{tab:table_just_alternative1}
\end{table}

\section{JAM Dynamical Model Test with Tangential Anisotropy \texorpdfstring{$\boldsymbol{\sigma_{\phi}/\sigma_{R}}$}{}}

\label{sec:appendix 2}
\begin{figure*}
    \centering
    \includegraphics[width=\linewidth]{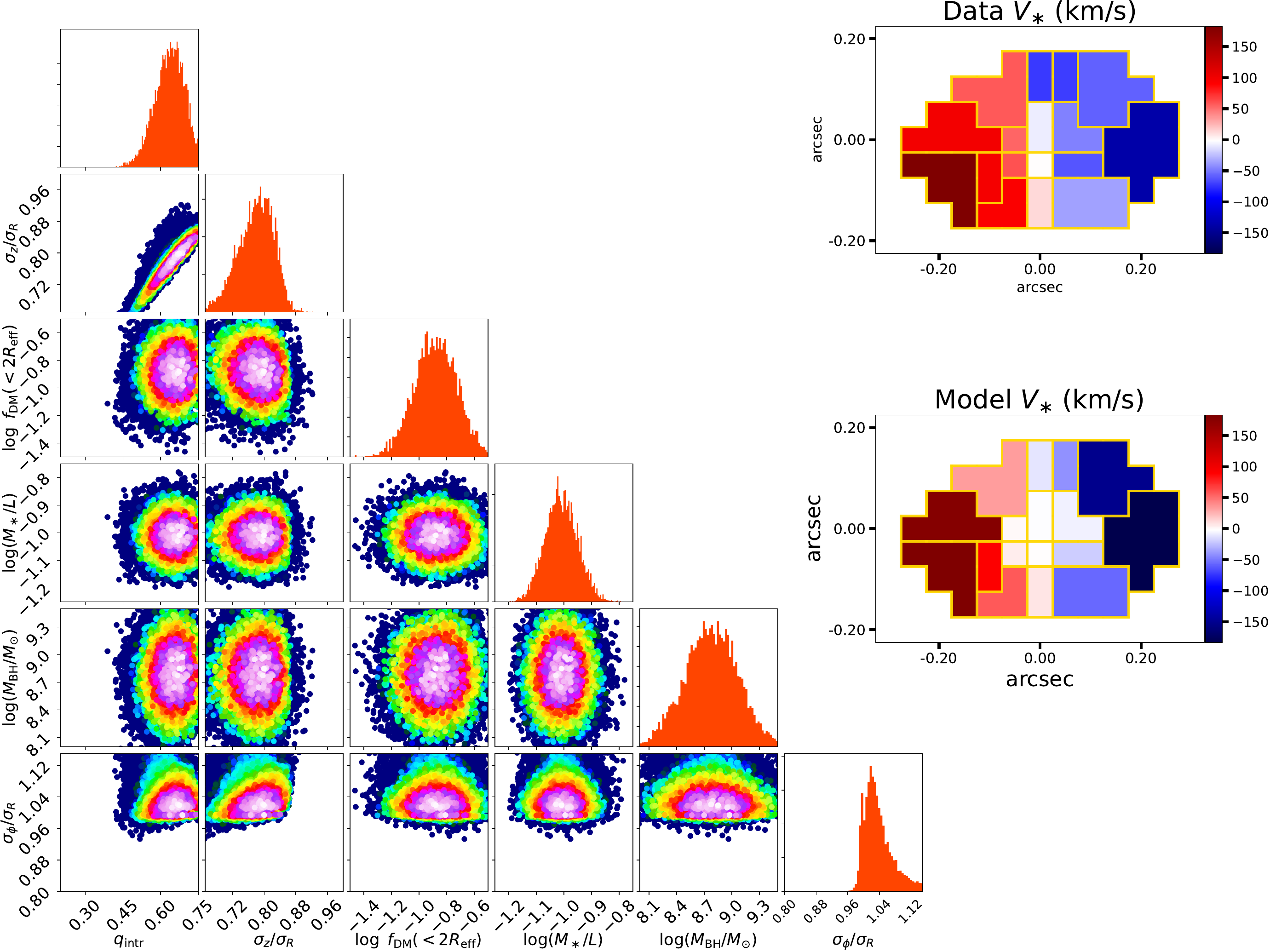}
    \caption{This figure shows the dynamical model \textit{Alternative-2} which fits the first LOSVD moment of the stellar kinematics, $V_{\ast}$. This model has similar characteristics to the \textit{Fiducial} model except that now we allow for a tangential anisotropy $\sigma_{\phi}/\sigma_{R}$ (with a Gaussian prior of mean = 1, standard deviation =0.2, this choice being motivated by the second panel of figure 2 from \citealt{sauron10}). We also impose Gaussian priors on the other five parameters, as follows: $q_{\rm intr}$ (mean: 0.62, standard deviation: 0.10), $\sigma_{\rm z}/\sigma_{\rm R}$ (mean: 0.846, standard deviation: 0.08), $\log f_{\rm DM} \left(<2R_{\rm eff}\right)$ (mean: -0.84, standard deviation: 0.16), $\log \left[\left(M_{\ast}/L \right)/\left(M_{\odot}/L_{\odot}\right) \right]$ (mean: -1.01, standard deviation: 0.06), $\log \left(M_{\rm BH}/M_{\odot}\right)$ (mean: 8.8, standard deviation: 0.3). These priors are motivated by the output results of the \textit{Fiducial} dynamical model. In this plot, we denote $\log \left(M_{\ast}/L\right) \equiv \log \left[ \left(M_{\ast}/L\right)/\left(M_{\odot}/L_{\odot}\right) \right]$.}

    \label{fig:tangential_anisotropy}
\end{figure*}

In this case, we use a dynamical model similar to our \textit{Fiducial} model: classical NFW Dark Matter Halo density profile, but this time we fit $V_{\ast}$ instead of $V_{\rm rms}$. We adopt Gaussian priors on all the parameters to be fitted, as explain in the caption of Fig. \ref{fig:tangential_anisotropy}. As this figure shows, \textit{Alternative-2} dynamical model constrains the tangential anisotropy to be $\sigma_{\phi}/\sigma_{\rm R} =1.01 \pm 0.05$. This quantity is important because it is used in de-projection calculations.

\begin{table}
\centering
\renewcommand{\arraystretch}{1.35} 
\setlength{\tabcolsep}{14pt}      
\captionsetup{justification=centering}
\caption{Best-fit parameters and uncertainties from \textit{JAM} using the dynamical model \textit{Alternative-2} (tangential anisotropy, NFW DM density profile). The Gaussian priors on each parameter are given in the caption of Fig. \ref{fig:tangential_anisotropy}}. 

\begin{tabular}{ c  c  c }
\hline
\textbf{parameter} & \textbf{Best-fit Value} \\ 
\hline
$q_{\rm intr} $ & $0.64 \pm 0.06$ \\ 
$\sigma_{z} / \sigma_{R}$ & $0.79 \pm 0.05$ \\ 
$\log f_{\rm DM} \left(<2R_{\rm eff}\right)$ & $-0.88 \pm 0.16$ \\
$\log \big{[}\left ( M_{\ast}/L \right )/\left(M_{\odot}/L_{\odot}\right) \big{]} $ & $-1.01 \pm 0.06$ \\ 
$\sigma_{\phi}/\sigma_{R}$ & $1.01 \pm 0.05$ \\ 
$\log \left(M_{\rm BH}/M_{\odot}\right)$ & $8.85 \pm 0.29$ \\
\hline
\end{tabular}

\label{tab:best_fit_Alternative-2}
\end{table}

\section{JAM Dynamical Model Test with a Gaussian Prior on the Intrinsic Axial Ratio}
\label{sec:appendix_prior}
\begin{figure*}
    \centering
    \includegraphics[width=\linewidth]{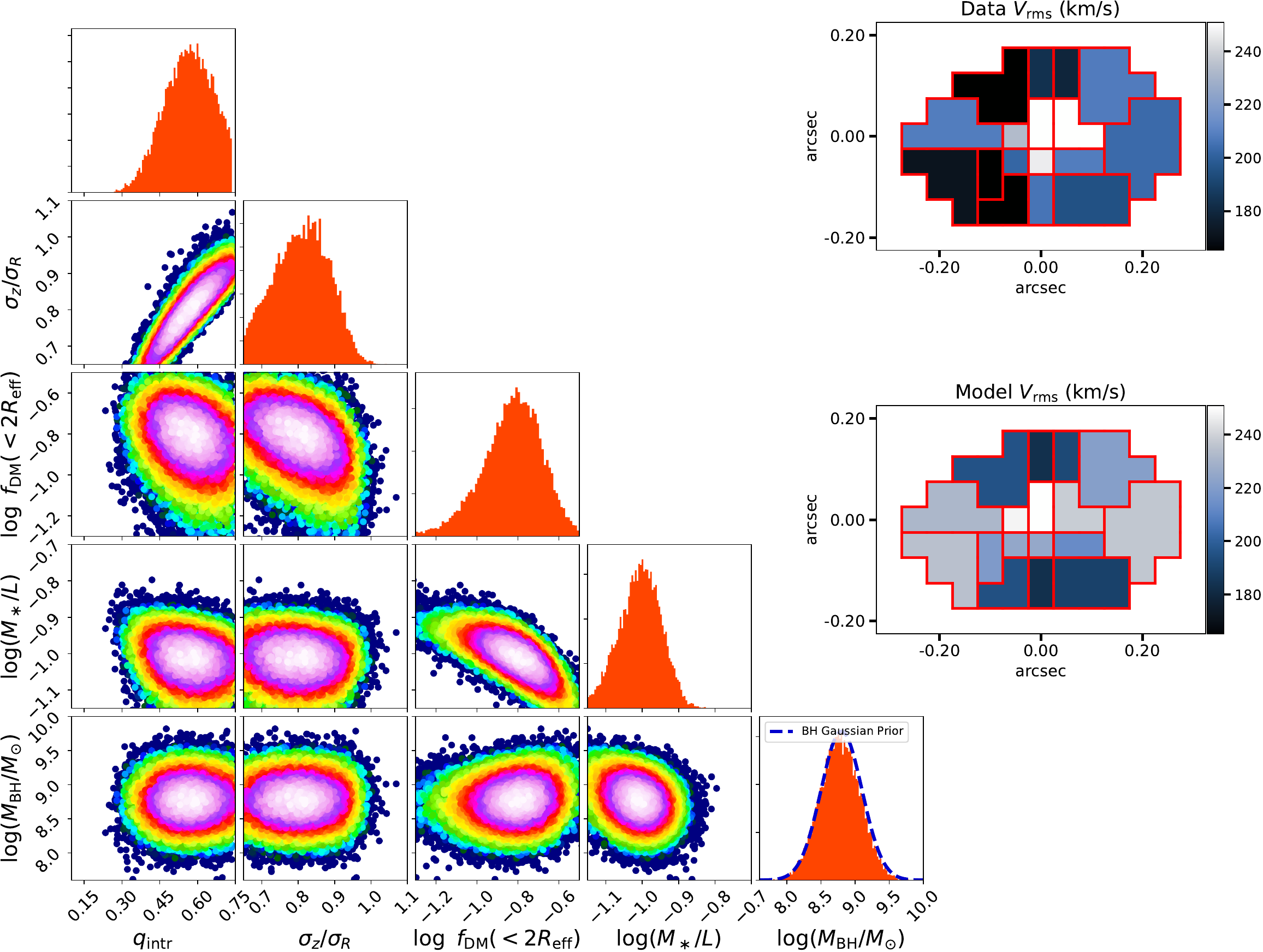}
    \caption{This figure shows the dynamical model \textit{Alternative-3} which has similar characteristics to \textit{Fiducial} except that now we force a Gaussian prior on $q_{\rm intr}$ (mean = 0.41, standard deviation = 0.18) allowing $\sigma_{\rm z}/\sigma_{\rm R}$, $\log \ f_{\rm DM} \left(<2R_{\rm eff}\right)$ and $\log \left[ \left(M_{\ast}/L\right)/\left(M_{\odot}/L_{\odot}\right)\right]$ as free parameters. In this plot, we denote $\log \left(M_{\ast}/L\right) \equiv \log \left[ \left(M_{\ast}/L\right)/\left(M_{\odot}/L_{\odot}\right) \right]$. This model has a reduced $\chi_{\rm r}^{2}=1.65$, which is larger than $\chi_{\rm r}^{2} =1.4$ of the \textit{Fiducial} dynamical model.}

    \label{fig:prior_on_q}
\end{figure*}

\begin{table}

\centering
\renewcommand{\arraystretch}{1.35} 
\setlength{\tabcolsep}{14pt}      
\captionsetup{justification=centering}
\caption{Best Fit parameters from \textit{JAM} using the dynamical model \textit{Alternative-3} (NFW DM density profile, Gaussians priors on $q_{\rm intr}$ and $\log M_{\rm BH}$). The Gaussian priors on each parameter are given in the caption of Fig. \ref{fig:prior_on_q}}. 
\begin{tabular}{ c  c  c }
\hline
\textbf{parameter} & \textbf{Best-fit Value} \\ 
\hline
$q_{\rm intr} $ & $0.57 \pm 0.09$ \\ 
$\sigma_{z} / \sigma_{R}$ & $0.81 \pm 0.09$ \\ 
$\log f_{\rm DM} \left(<2R_{\rm eff}\right)$ & $-0.82 \pm 0.14$ \\
$\log \big{[}\left ( M_{\ast}/L \right )/\left(M_{\odot}/L_{\odot}\right) \big{]} $ & $-1.01 \pm 0.06$ \\ 
$\log \left(M_{\rm BH/M_{\odot}} \right)$ & $8.79 \pm 0.29$ \\ 

\hline
\end{tabular}

\label{tab:best-fit-alternative3}
\end{table}

This dynamical model fits $V_{\rm rms}$ and it is similar to the \textit{Fiducial} model, the only difference being the fact that now we introduce a Gaussian prior on $q_{\rm intr}$ (in addition to the already existing prior on the central BH mass). The reason for this test is to probe to what extent the predictions on $q_{\rm intr}$ that are purely based on our data alone can win against a physically motivated prior. The results of \textit{Alternative-3} dynamical model fitting are given in \autoref{tab:best-fit-alternative3}.

\section{Dark Matter Content - Comparison with the \textsc{TNG-50} simulations}
\label{sec:appendix_dm}

\begin{figure*}
    \centering
          
      
        \label{fig:TNG100_and300}

            \includegraphics[width=\linewidth]{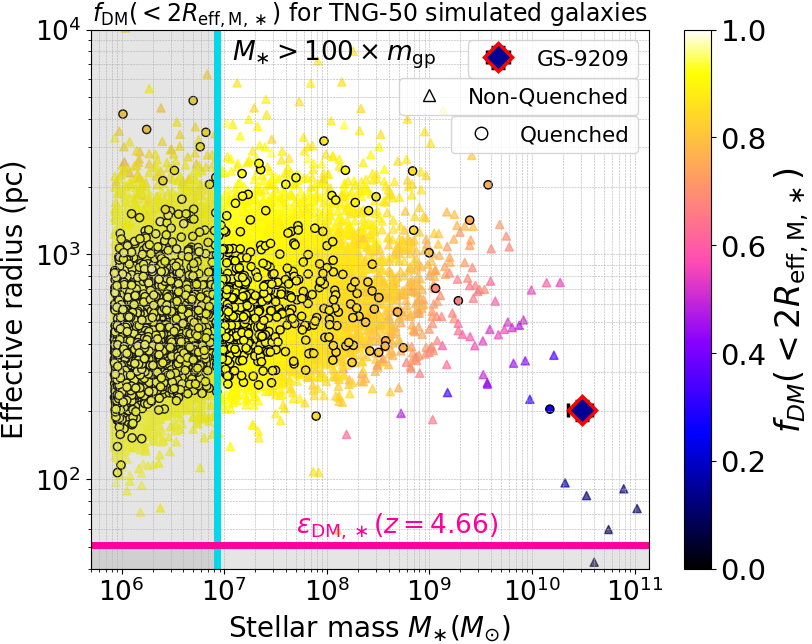}
        
     \caption{Dark matter fraction within $2 \ R_{\rm eff,M,\ast}$ of simulated galaxies at $z \sim 4.66$ in \textit{TNG50-1} run. We observe that as we go towards smaller effective radii and larger stellar masses, $f_{\rm DM}$ decreases. There is good agreement between the value obtained from our \textit{JAM} dynamical modelling algorithm $f_{\rm DM} \left(<2R_{\rm eff,M,\ast}\right) = 14.5^{+6.0}_{-4.2} \%$ and the predictions given by the \textit{TNG-50} simulation. We remark a similar behaviour for the variation of $f_{\rm DM} \left(<2R_{\rm eff, M,\ast} \right)$ across the mass-size plane for the quiescent versus non-quiescent simulated galaxies. The quiescence criterion is $\rm sSFR < 0.2/t_{H} \left(z =4.66\right)$. The light blue line is the threshold below which the stellar masses of the simulated galaxies are smaller than $100\times m_{\rm gp}$ (where $m_{\rm gp}$ is the mass of a gas particle). The pink horizontal line is the physical Plummer-equivalent gravitational softening parameter $\epsilon_{\rm DM,\ast} \left(z =4.66\right)$ (for collisionless particles i.e. dark matter and stars). This is calculated as $\epsilon_{\rm DM,\ast} \left(\rm z\right) = \epsilon_{\rm DM,\ast}^{\rm z=0} / \left(\rm 1+z\right)$ with $\epsilon_{\rm DM,\ast}^{\rm z=0}=290 \ \rm pc$ for \textit{TNG50-1}. The plotted simulations data were smoothed using the 2-dimensional Locally-Weighted Regression (\textit{LOESS}; \citealt{loess2d}) to ensure that each colour-coded data point shows the local average trend and to remove local outliers.}
     \label{fig:ALL4_TNG}
\end{figure*}

\begin{table}
\centering
\renewcommand{\arraystretch}{1.45} 
\setlength{\tabcolsep}{12pt}      
\captionsetup{justification=centering}
\caption{Convergence radii (using the criterion from \citealt{ludlow+18}) for various simulations that we considered for analysing their predictions for the dark matter content of GS-9209 at $z=4.66$. 
This table justifies our choice of \textsc{TNG50-1} in our analysis: this box has a similar spatial resolution to GS-9209 effective radius and the box size is still reasonably large (50 cMpc/h) to probe a reasonable cosmic volume at $z=4.66$.}

\begin{tabular}{ c  c}
\hline
Simulation box & $\rm R_{\rm conv}$ (pc) \\ 
\hline

\textsc{TNG50-1} & 233 \\
\textsc{TNG100-1} & 591 \\
\textsc{Illustris-1} & 568 \\
\textsc{Eagle} - 100 cMpc box & 646 \\
\textsc{Eagle} - 25 cMpc box & 323 \\
\textsc{Magneticum} - uhr resolution & 1195 \\
\textsc{Magneticum} - xhr resolution & 447 \\
\textsc{COLIBRE} - L050m5 box & 301 \\
\textsc{COLIBRE} - L200m6 box & 601 \\
\textsc{SIMBA}  & 1406 \\
\textsc{FLAMINGO} - best resolution & 2700 \\

\hline

\end{tabular}

\label{tab:convegences}
\end{table}

Based on their study of \textit{EAGLE} hydrodynamical simulations \citep{eagle}, \citet{ludlow+18} propose the following criterion for simulated galaxies size convergence:

\begin{equation}
R_{\rm gal} \geq R_{\rm conv} \equiv 0.055 \ \rm L\left(\rm z\right),
\end{equation}

\noindent with $L\left(\rm z\right)$ being the mean spacing between dark matter particles, calculated using the equation $L\left(\rm z=0\right) = L_{\rm box}/N_{\rm DM}^{1/3}$ and $L\left(z\right) = L \left(\rm z=0\right) /\left(1+z\right)$. The values for the threshold convergence radius for a number of \textit{TNG} simulation boxes \citep{nelson2019TNG}, as well as for the \textit{EAGLE} \citep{eagle}, \textit{Illustris-1} \citep{pillepich+2018,springel+2018}, \textit{SIMBA} \citep{simba_dave2019}, \textit{FLAMINGO} \citep{flamingo}, \textit{Magneticum} \citep{relight}, and \textsc{COLIBRE} \citep{colibre_2025} cosmological simulations, are given in \autoref{tab:convegences}. For GS-9209, the effective light radius is $R_{\rm eff} \approx 223 \pm 20 \, \rm pc$, which is below the convergence radius calculated using this criterion, for all simulations considered except for \textit{TNG50-1} (which has a convergence radius comparable to GS-9209 effective radius). This means that all the other simulations cannot accurately probe the existence and number densities of extremely compact, high-redshift MQGs, because the size convergence is uncertain for simulated galaxies which have $R_{\rm eff}<R_{\rm conv}$. We will therefore focus on analysing the MQGs produced within the \textit{TNG50-1} simulation box.

We compare $f_{\rm DM} \left(<2R_{\rm eff} \right)$ of GS-9209 with the values in the cases of simulated galaxies in the \textit{TNG50-1} box. 
However, these simulations compute the dark matter fraction within twice the stellar half mass radius $\rm 2R_{\rm eff,M,\ast}$ (twice the radius enclosing half of the total stellar mass within a given subhalo). \textsc{PySersic} computed the projected effective light radius obtained from the galaxy's photometry data in NIRCam/F200W band, $R_{\rm eff,2D}$. The first step is to use \textit{JAM} dynamical modelling to compute the 3-dimensional intrinsic half light radius $\rm R_{\rm eff,lum,3D}$ (the radius of a sphere enclosing half of the total light emitted over the entire volume of the galaxy) of GS-9209. This is done by using an entirely analogous procedure to \textsc{jam\_mge\_half\_light\_radius}. The latter computes the radius of a circle enclosing half of the light within the surface brightness Gaussian surfaces (characterised by $I_{j}$, $\sigma_{j}$ and $q_{j}^{\prime}$ using the same notations as in Section \ref{sec:mgefit}) from the \textit{MGE} parametrisation of the galaxy's S\'{e}rsic profile fitted by \textsc{PySersic} to the original image. Instead, we need to convert these surface brightness Gaussians to luminosity density Gaussians as we did in Section \ref{sec: kappa_rot}. The intrinsic luminosity density Gaussians are characterised by $\nu_{j}$, $\sigma_{j}$ and $q_{j}$ (with the same notations as in Equation \ref{eq:density_MGE}). 
The integrated luminosity of a Gaussian is given by (equation 13 from \citealt{cappellari2008jam}): 

\begin{equation}
L_{j} = \nu_{j} q_{j} \times \left(\sigma_{j}\sqrt{2\pi}\right)^{3} .
\end{equation}

\noindent The total luminosity of all the Gaussians is the sum of $L_{\rm j}$ values. This allows us to compute the radial distance at which Gaussians within that aperture only add up to half the total luminosity. We obtain that, for GS-9209, $R_{\rm eff,lum,3D} = 0.91 \ R_{\rm eff, 2D}$ where $R_{\rm eff,2D}$ is the effective light radius from the F200W image. This is in good agreement with the findings highlighted in the top-left panel of figure 4 from \citet{van_de_ven2022} who determine similar values for the ratio between the 2-dimensional projected effective light radius and the 3-dimensional intrinsic counterpart. Because GS-9209 is a massive quiescent galaxy, it is reasonable to assume that its gas mass is negligible. Future studies using the G395H / F270LP spectrum  from NIRSpec will allow us to quantify neutral gas in the galaxy's interstellar medium, whereas ALMA could determine the amount of cold gas (or at least impose a stringent upper limit for the cold gas mass in GS-9209). Studies such as \citet{jan_net0} showed that for GS-10578, a MQG at $z\sim3$, the CO-derived cold gas mass is less than 3\% of its $M_{\ast}$ whereas \citet{deugenio+2024} determine that the neutral gas mass of GS-10578 is about $10^{8} \ \rm M_{\odot}$ (derived from NaD absorption). The upper limits found by \citet{suzuki_lowgas} on the gas fractions of 5 MQGs at $3.5<z<4$ and with $10.5<\log \left(M_{\ast}/M_{\odot}\right)<11.0$ are less stringent ($f_{\rm gas} \leq 20\%$) but consistent with the general picture that high redshift MQGs are highly likely to contain very little cold gas. For these reasons, we can consider that stars are the only luminous component in our model of GS-9209. Assuming no spatial variation of the stellar mass-to-light ratio, it means that $R_{\rm eff,lum,3D} = R_{\rm eff,M,\ast} = 0.91 \ \rm R_{\rm eff} = 203 \ \rm pc$. This result is in agreement with the findings from figure 8 of \citet{degraaff+21} based on $z \sim 0.1$ galaxies from \textit{EAGLE} simulations, who obtain that for the most compact galaxies in their sample, their photometry-based effective radii are typically slightly larger than the mass effective radii, but the difference is less than 0.1 dex. 

In order to compute $f_{\rm DM}$ within $2 \ R_{\rm eff,M,\ast}$ we use the method outlined below. We chose the snapshot that is closest to the redshift of GS-9209, which has the ID ``18'' and redshift $z_{\rm snap} \approx 4.66$. We use the provided group catalogues to extract all the galaxy or halo properties needed in the subsequent analysis. Specifically, all the quantities we need at this point reside in the \textit{Subhalos} data repositories (these are retrieved using the \textit{Subfind} algorithm modified to account for the properties and behaviour of the galaxies' baryonic component). For our analysis we only need the quantities that can be queried using the keywords: \textsc{SubhaloMassInRadType}, \textsc{SubhaloHalfmassRadType}, \textsc{SubhaloMassType} and \textsc{SubhaloSFR}. \textsc{SubhaloFlag} is used to eliminate possible artefacts from our sample. This is particularly important for our analysis because such contaminating subhalos are typically compact and baryon dominated (although usually more than 3 orders of magnitude less massive than GS-9209). They are not individual galaxies but clumps within genuine galaxies. The first keyword is used to extract, from the group catalogue, the masses for each component of a galaxy (gas, DM, stars + winds, BH) within $2 \ R_{\rm eff,M,\ast}$, and hence to compute the DM fraction within this aperture $f_{\rm DM} \left( < 2R_{\rm eff,M,\ast}\right) =  M_{\rm DM} \left (< 2R_{\rm eff,M,\ast}\right ) / M_{\rm tot}  \left (< 2R_{\rm eff,M,\ast}\right )$. The total mass is obtained by adding up all the particles of all types within $<2 R_{\rm eff,M,\ast}$ from the subhalo's centre of mass. The second keyword gives the stellar half mass radii of the galaxies in each simulation. The units in the catalogue are ckpc/h so in order to convert to pkpc, we take h = 0.6774 (this value is adopted by TNG simulations) and we divide by $\left(1+z_\mathrm{snap} \right)$. The \textsc{SubhaloMassType} keyword gives the masses of each component of a galaxy that is gravitationally bound to that subhalo. In our case we are only interested in the stellar mass.

The results of $f_{\rm DM} \left (<2 \ R_{\rm eff,M,\ast} \right)$ variation across the $M_{\ast} - R_{\rm eff,M,\ast} $ are shown in \autoref{fig:ALL4_TNG}. This plot shows that the low dark matter content of GS-9209 is in good agreement with the values obtained in the case of its simulated counterparts. Furthermore, it is expected that we find a low amount of dark matter within two effective stellar half-mass radii for GS-9209 given its extremely compact size. Analysing the plot, we find that the \textit{TNG50-1} simulation cannot reproduce massive quiescent galaxies with similar stellar masses and effective radii to GS-9209 at $z=4.66$. 

 
Overall, GS-9209 occupies a notably different position on the mass-size plane than the majority of the population of quiescent galaxies in this simulation (marked as circles). By carefully analysing the high mass and low-effective radius end of the diagram ($M_{\ast}>10^{10} \, M_{\odot}$; $R_{\rm eff}<300 \, \rm pc$), we notice that there is only one compact MQG similar to GS-9209. This object has a slightly higher dark matter fraction of than the value we infer for GS-9209, but it is still unambiguously baryon dominated. 
This allows us to estimate the number density of compact massive quiescent galaxies (c-MQGs) like GS-9209 (as predicted by simulations): $n_{\rm c-MQG} \sim \left(\mathrm{50/h \ Mpc }\right)^{-3}=2.4 \times 10^{-6} \rm Mpc^{-3}$. This is approximately one order of magnitude below the calculated number densities values that were reported in various $z = 3-5$ recent MQGs observational studies (e.g \citealt{glazebrook_jekyll_2017,carnall_CEERS_abundance2023,carnall2023excels,Valentino_2023_nquench,minjungpark2024_outflows}). This result can be explained in two different ways, one of them being the claim that even highly performant cosmological full hydro-dynamical simulations with the best spatial resolution might be unable to fully capture the physical processes that govern the formation and evolution of c-MQGs in the early Universe. However, it is also likely that these simulations do not accurately reproduce compact galaxies, primarily because of size convergence caveats (for \textit{TNG50-1}, $R_{\rm conv}$ is approximately equal to the effective light radius of GS-9209). Even if \textit{TNG50-1} can theoretically probe the regime of compact MQGs, it still struggles to reproduce the existence of objects similar to GS-9209, highlighting the need for building cosmological simulations that have both a large box volume ($\geq100 \ \rm cMpc/h$) and good enough spatial resolution ($<200 \ \rm pc$; as defined based on the convergence criterion of \citealt{ludlow+18}). This would be required in order to test whether the disagreement between the observed number densities of high-$z$ MQGs and the predictions of simulations is an issue related to number statistics and computational limitations of the simulations or, alternatively, if the physical mechanisms implemented need to be revised.  

\section{JAM recovery tests of galaxy's dynamical parameters}

\label{sec: appendix E}

In this section we present some example recovery tests we conducted on our fiducial dynamical model used in the main text. For each of these runs in particular, we use \textsc{jam\_axi\_proj} routine to construct a model galaxy data based on a set of 4 parameters: the intrinsic axial ratio, $q_{\rm intr}$, radial anisotropy ratio $\sigma_{z}/\sigma_{R}$, the stellar mass to light ratio $\log \left[\left((M/L \right)_{\rm tot}/\left(M_{\odot}/L_{\odot}\right)\right]$ and the logarithmic dark matter fraction within one effective radius, $\log f_{\rm DM} \left(<2R_{\rm eff}\right)$. 
We have chosen that in this case we will not use the Gaussian prior on the central BH mass and instead just fix it to $M_{\rm BH} =10^{8.8} M_{\odot}$ (for speeding up the computations). We perform the tests as follows:  


\begin{itemize}
    \item We start from the values of the parameters obtained for GS-9209 (Table \ref{tab:table_fiducial_alternative}). For each parameter, we run one recovery test.
    \item We vary only one parameter at a time. For instance, for the $q_{\rm intr}$ recovery test, all the other parameters besides $q_{\rm intr}$ are taken to be the values found in \autoref{tab:table_fiducial_alternative} (and the BH mass is taken as $10^{8.8} M_{\odot}$). We then assign an arbitrary value for the parameter we want to test, sufficiently different from the value we had in the case of GS-9209 (e.g. $q_{\rm intr}=0.3$ in our example case that we described). 
    \item We keep doing one such test per parameter per dynamical model. However we do not test for $M_{\rm BH}$ since this value cannot be constrained given the kinematics resolution we have. We test for all the other parameters instead. 
    For the light distribution map we use the same values as obtained from the \textsc{MgeFit} in the case of GS-9209. We also use the same distance, same Voronoi bins rotated coordinates, same pixel scale as we did for GS-9209 (as keywords for \textsc{jam\_axi\_proj}).
    \item After initialising the set of parameters for a test, the next thing to do is to build the model galaxy kinematics using \textsc{jam\_axi\_proj} routine with no input data or error maps. We add the noise after creating the model galaxy using the same noise map as for GS-9209. We pass all of these to the $\chi^{2}$ \textit{JAM} minimisation routine and we run \textsc{adamet} with $10^{4}$ steps (instead of $10^{5}$ as we did in the main text dynamical modelling). This is done in order to illustrate the validity of the recovery tests, of the models we used and to allow the computation of more dynamical models tested in a short time. Nevertheless, even in the case of the original dynamical models, the convergence is reached quickly enough that if we use $10^{4}$ steps in \textsc{adamet} instead of $10^{5}$, this does not produce a noticeable difference in terms of the retrieved best-fit parameters from the posterior distribution.
\end{itemize}



\begin{figure*}
    \centering
    \includegraphics[width=\linewidth,height=9.75cm]{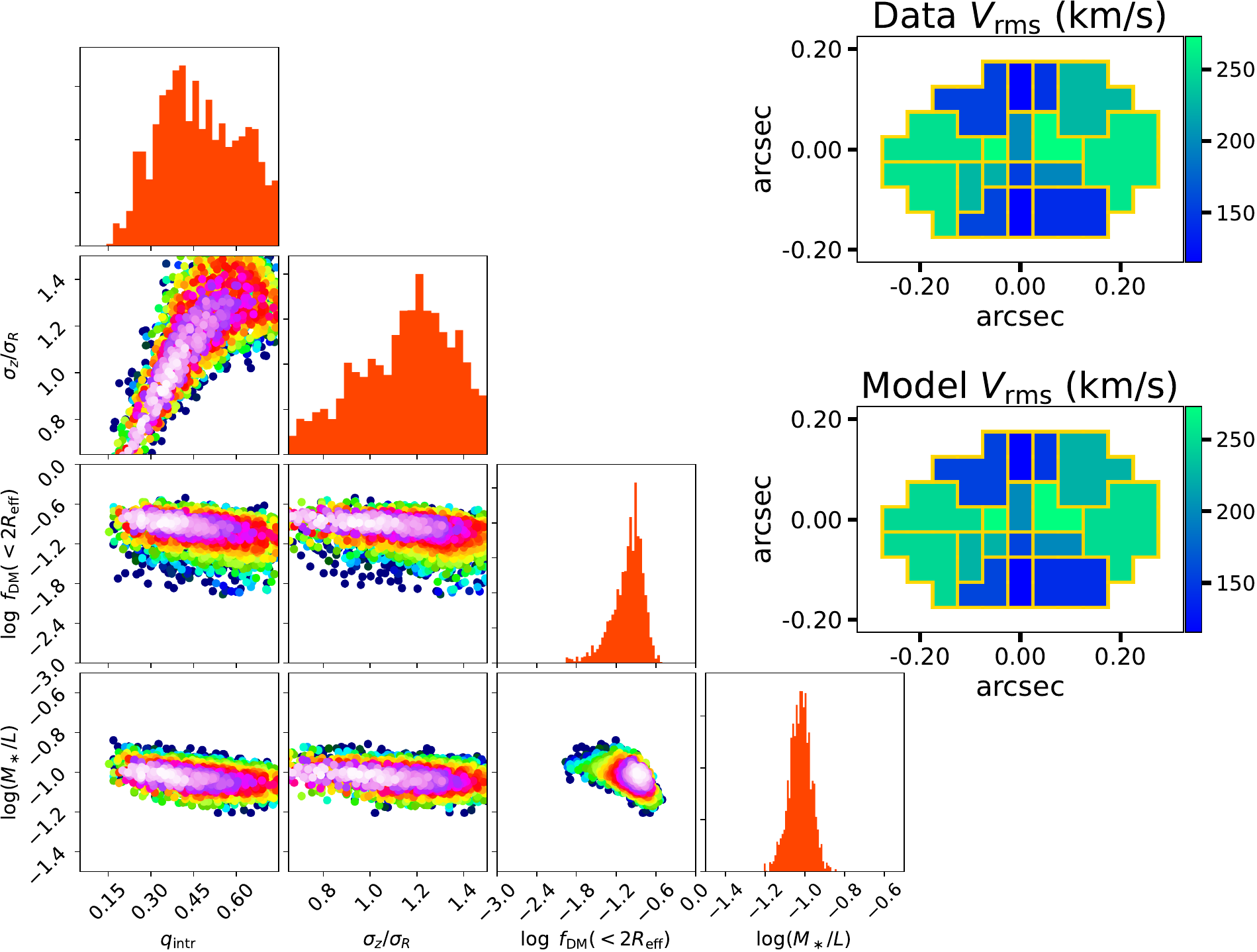}
    \caption{Fiducial recovery test for $q_{\rm intr}$. The tested input value is $q_{\rm test,in} = 0.3$ and the test output is $q_{\rm test,out} = 0.35 \pm 0.12$. We have $\left(\sigma_{z}/\sigma_{R}\right)_{\rm test,out}= 0.94 \pm 0.22$ (consistent with the input value of 0.85), $\log \left (M_{\ast}/L \right)_{\rm test,out} = -1.02 \pm 0.05$ (in agreement with the input value of -1.01) and $ \left(\log \ f_{\rm DM,<2R_{\rm eff}} \right)_{\rm test,out} = 0.88 \pm 0.22$ (in agreement with the input value of -0.84). In this plot, we denote $\log \left(M_{\ast}/L\right) \equiv \log \left[ \left(M_{\ast}/L\right)/\left(M_{\odot}/L_{\odot}\right) \right]$.}

    \label{fig:rec_2}
\end{figure*}



\begin{figure*}
    \centering
    \includegraphics[width=\linewidth,height=9.75cm]{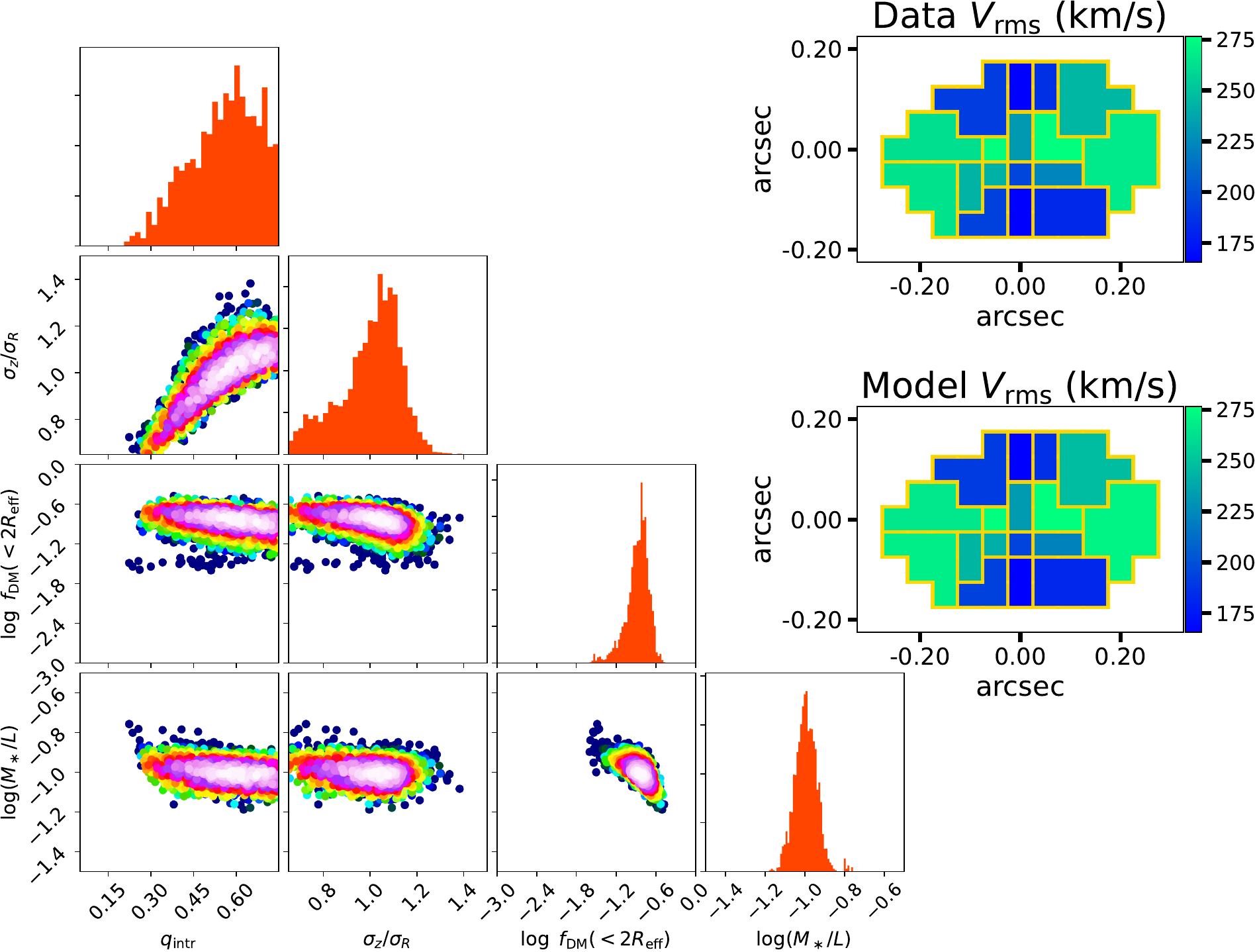}
    \caption{Fiducial recovery test for $\sigma_{z}/\sigma_{R}$. The tested input value is $\left ( \sigma_{z}/\sigma_{R}\right)_{\rm test,in} = 1.05$ and the test output is $\left(\sigma_{z}/\sigma_{R} \right)_{\rm test,out} = 1.00 \pm 0.13$. We have $q_{\rm test,out}= 0.55 \pm 0.13$ (in agreement with the input value of 0.62), $\log \left (M_{\ast}/L \right)_{\rm test,out} = -1.00 \pm 0.05$ (in agreement with the input value of -1.01) and $ \left(\log \ f_{\rm DM,<2R_{\rm eff}} \right)_{\rm test,out} = -0.80 \pm 0.16$ (in agreement with the input value of -0.84). In this plot, we denote $\log \left(M_{\ast}/L\right) \equiv \log \left[ \left(M_{\ast}/L\right)/\left(M_{\odot}/L_{\odot}\right) \right]$.}

    \label{fig:rec_4}
\end{figure*}



\begin{figure*}
    \centering
    \includegraphics[width=\linewidth,height=9.75cm]{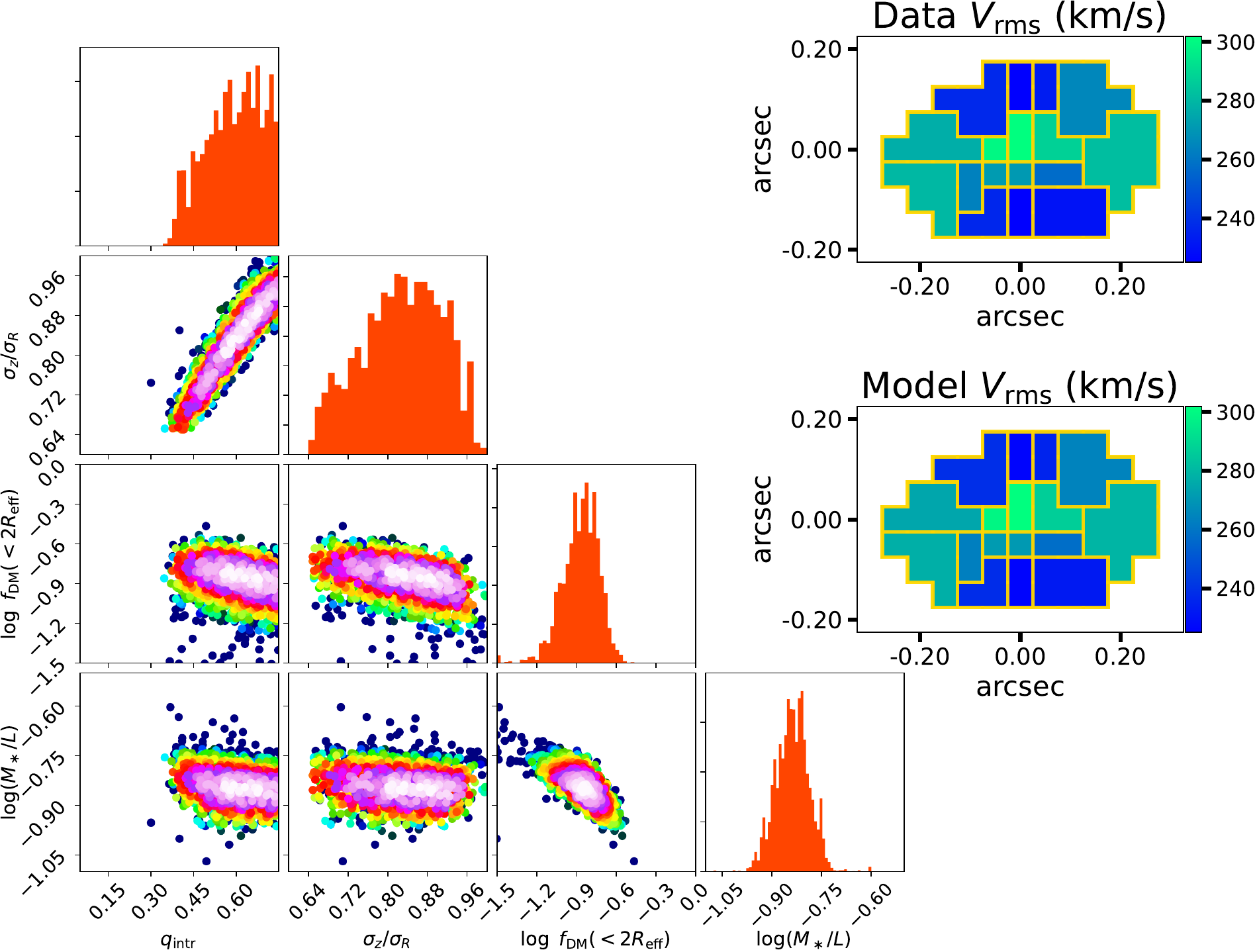}
    \caption{Fiducial recovery test for $\log \left(M_{\ast}/L \right)$. The tested input value is $\log \left (M_{\ast}/L \right)_{\rm test,in} = -0.85$ and the test output is $\log \left(M_{\ast}/L \right)_{\rm test,out} = -0.86 \pm 0.04$. We have $\left(\sigma_{z}/\sigma_{R}\right)_{\rm test,out}= 0.89 \pm 0.09$ (in agreement with the input value of 0.85), $q_{\rm test,out} = 0.66 \pm 0.11$ (in agreement with the input value of 0.62) and $\log \left(f_{\rm DM, <2R_{\rm eff}} \right)_{\rm test,out} = -0.85 \pm 0.14$ (in agreement with the input value of -0.84). In this plot, we denote $\log \left(M_{\ast}/L\right) \equiv \log \left[ \left(M_{\ast}/L\right)/\left(M_{\odot}/L_{\odot}\right) \right]$.}

    \label{fig:rec_6}
\end{figure*}

\begin{figure*}
    \centering
    \includegraphics[width=\linewidth,height=9.75cm]{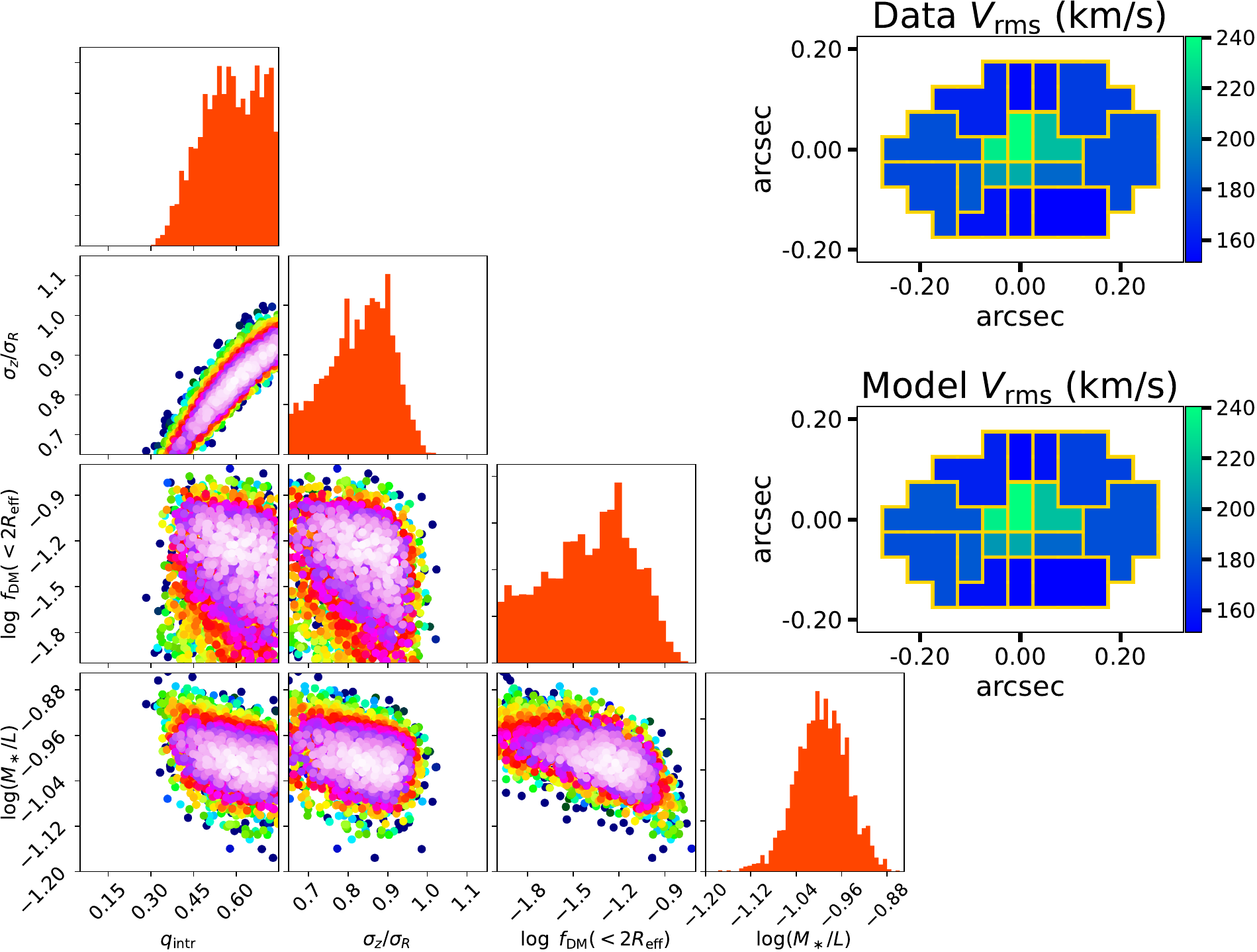}
    \caption{First fiducial model recovery test for $\log  f_{\rm DM} \left(<2R_{\rm eff}\right)$. The tested input value is $\left (\log  f_{\rm DM,<2R_{\rm eff}} \right)_{\rm test,in} = -1.25$ and the test output is $\left (\log  f_{\rm DM,<2R_{\rm eff}} \right)_{\rm test,out} = -1.22 \pm 0.32$. We have $\left(\sigma_{z}/\sigma_{R}\right)_{\rm test,out}= 0.835 \pm 0.084$ (in agreement with the input value of 0.85), $q_{\rm test,out}= 0.59 \pm 0.11$ (in agreement with the input value of 0.62) and $\log \left (M_{\ast}/L \right)_{\rm test,out} =-1.008 \pm 0.044$ (in agreement with the input value of -1.01). In this plot, we denote $\log \left(M_{\ast}/L\right) \equiv \log \left[ \left(M_{\ast}/L\right)/\left(M_{\odot}/L_{\odot}\right) \right]$.}

    \label{fig:rec_7}
\end{figure*}

\bsp	
\label{lastpage}
\end{document}